\documentclass[twocolumn]{aastex62}

\date {\today}

\shorttitle{Field Structure of Dense Cores}
\shortauthors{Auddy et al.}

\begin{document}

\title{Magnetic Field Structure of Dense Cores using Spectroscopic Methods }

\correspondingauthor{Sayantan Auddy}
\email{sauddy3@uwo.ca}

\author{Sayantan Auddy}
\affil{Harvard-Smithsonian Center for Astrophysics, 60 Garden Street, Cambridge, MA 02138, USA}
\affiliation{Department of Physics and Astronomy, The University of Western Ontario, London, ON N6A 3K7, Canada}

\author{ Philip C. Myers}
\affiliation{Harvard-Smithsonian Center for Astrophysics, 60 Garden Street, Cambridge, MA 02138, USA} 

\author{Shantanu Basu}
\affiliation{Department of Physics and Astronomy, The University of Western Ontario, London, ON N6A 3K7, Canada}
\author{Jorma Harju}
\affiliation{Max-Planck-Institute for Extraterrestrial Physics (MPE), Giessenbachstr. 1, 85748 Garching, Germany}
\affiliation{Department of Physics, P.O. Box 64, 00014 University of Helsinki, Finland}

\author{Jaime E. Pineda}
\affiliation{Max-Planck-Institute for Extraterrestrial Physics (MPE), Giessenbachstr. 1, 85748 Garching, Germany}

\author{Rachel K. Friesen}
\affiliation{National Radio Astronomy Observatory, Charlottesville, VA 22903, USA}




\begin{abstract}
We develop a new ``core field structure'' (CFS) model to predict the magnetic field strength and magnetic field fluctuation profile of dense cores using gas kinematics. We use spatially resolved observations of the nonthermal velocity dispersion from the Green Bank Ammonia survey along with column density maps from SCUBA-2 to estimate the magnetic field strength across seven dense cores located in the L1688 region of Ophiuchus. The CFS model predicts the profile of the relative field fluctuation, which is related to the observable dispersion in direction of the polarization vectors. Within the context of our model we find that all the cores have a transcritical mass-to-flux ratio.
\end{abstract}





\section{Introduction} \label{sec:intro}
Stars form in dense cores embedded within interstellar molecular clouds \citep{lad93,wil00,and09}. Dense cores are well studied observationally from molecular spectral line emission \citep{mye83,ben89,jij99}, infrared absorption \citep{tei05,lad07,mac17} and submillimeter dust emission \citep{war94,kir05,mar16}.

Cores may form in multiple ways including fragmentation of over-dense regions that are typically filaments and sheets \citep{bas09,bas09a} within turbulent magnetized clouds. Depending on the ambient initial conditions they can form either as a result of spontaneous gravitational contraction \citep{jea29,lar85,lar03} or by rapid fragmentation due to preexisting turbulence \citep{pad97a,kle01,gam03}. Another scenario is the formation of cores in magnetically supported clouds due to quasistatic ambipolar diffusion i.e., gravitationally induced drift of the neutral species with respect to ions  \citep{mes56,mou79,shu87}. However, a more recent view is that both supersonic turbulence and gravitationally driven ambipolar diffusion are significant in the process of core formation \citep[e.g.,][]{nak05,kud11,kud14,che14,aud18}. 

Dense cores often have nonthermal contributions to line width that are small compared to the thermal values \citep{ryd77,mye83a,cas02}. These observations imply a transition from a primarily nonthermal line width in low density molecular cloud envelopes to a nearly thermal line width within dense cores. This is termed as ``a transition to velocity coherence" \citep{god98}. A  sharp transition between the coherent core and the dense turbulent gas surrounding the B5 region in Perseus was found using NH$_{\rm {3}}$ observations from the Green Bank Telescope (GBT) by \cite {pin10}. It has been suggested that this transition arises from damping and reflection of MHD waves \citep{pin12}.

An important question is whether a transition from magnetic support of low density regions to gravitational collapse of dense regions is physically related to the transition to coherence. Furthermore, how is the magnetic field strength affecting the nonthermal line width in the low density region, and is this related to the velocity transition? If so, can one estimate the magnetic field strength and its radial variation across a dense core using such observations?  

Accurate measurement of the magnetic field is one of the challenges of observational astrophysics. Several methods exist that probe the magnetic field in the interstellar medium, such as Zeeman detection \citep[e.g.,][]{cru99}, dust polarization \citep{hoa08} and Faraday rotation \citep{wol04}. While each method has its own limitations \citep{cru12}, sensitive observations of dust polarization can describe the structure of the plane-of-sky magnetic field and can estimate its strength. According to the dust alignment theory \citep{and15}, the elongated interstellar dust grains tend to align with their minor axis parallel to the magnetic field. Dust polarization observations from thermal emission or extinction of background starlight provide a unique way to probe the magnetic field morphology in the ISM, including collapsing cores in molecular clouds. 

In addition to getting the field morphology, there are various methods to estimate the magnetic field strength. 
One of the popular techniques is the Davis-Chandrasekhar-Fermi (DCF) method \citep{dav51,cha53} that estimates the field strength using measurements of the field dispersion (about the mean field direction), gas density, and one-dimensional nonthermal velocity dispersion. Dust polarization, however, can be weak in the centers of dense cores where the dust grains are well shielded from the radiative torques necessary to move the grains into alignment with the magnetic field \citep[e.g., see][]{laz07}. 

Recently, \cite{mye18} have extended the spherical flux-freezing models of \cite{mes66} and \cite{mes67} to spheroidal geometry, allowing quantitative estimates of the magnetic field structure in a variety of spheroidal shapes and orientations. In these models the magnetic field energy in the spheroid is weaker than its gravitational energy, allowing gravitational contraction, which drags field lines inward. However the spheroid magnetic field energy is stronger than its turbulent energy, allowing the field lines to have an ordered ``hourglass" structure.  These models are useful to test clouds, cores, and filaments that show ordered polarization for the prevalence of flux freezing. They also allow an estimate of the magnetic field structure when the underlying density structure is sufficiently simple and well known.  

The present paper is complementary to \cite{mye18} since it relies on many of the same assumptions of flux freezing in a centrally condensed star-forming structure.  However it is more specific to dense cores having subsonic line widths.  It also relies on additional information, i.e., maps of nonthermal line widths, and on the additional assumption that the nonthermal line widths are due to Alfv{\'e}nic fluctuations in the magnetic field lines, as in the original studies of DCF.

In this paper we predict magnetic field structure by analyzing new NH$_{3}$ observations of multiple cores in the L1688 region in the Ophiuchus molecular cloud. Most of the cores show a sharp transition to coherence with a nearly subsonic nonthermal velocity dispersion in the inner region. 
We propose a new $``$Core Field Structure$"$ (CFS) model of estimating the amplitude of magnetic field fluctuations. It incorporates detailed maps from the Green Bank Ammonia Survey (GAS) of the nonthermal line width profiles across a core. 
The paper is organized in the following manner. The observations of the gas kinematics and the column density are reported in Section 2. In Section 3 we introduce the CFS model and the inferred magnetic field profile. In Section 4 we discuss the limitations of the model. We highlight some of the important conclusions in Section 5.

\section{Observations}

The first data release paper from the GBT survey \citep{fri17} included detailed NH$_{3}$ maps of the gas kinematics (velocity dispersion, $\sigma_{v}$ and gas kinetic temperature, $T_{K}$) of four regions in the Gould Belt: B18 in Taurus, NGC 1333 in Perseus, L1688 in Ophiuchus, and Orion A North in Orion. The emission from the NH$\rm _{3}$ $(J,K)$ $=$ $(1,1)$ and $(2,2)$ inversion lines in the L1688 region of the Ophiuchus cores were obtained using the $100$ m Green Bank Telescope. The observations were done using in-band frequency switching with a frequency throw of $4.11$ MHz, using the GBT K-band (upper) receiver and the GBT spectrometer at the front and back end, respectively. L1688 is located centrally in the Ophiuchus molecular cloud. It is a concentrated dense hub (with numerous dense gas cores) spanning approximately $ 1-2$ pc in radius with a mass of $ 2 \times 10^3\, \rm {M_{\odot }}$ \citep{lor89}.  L1688 has over 300 young stellar objects \citep{wil08} and contains regions of high visual extinction, with A$_{\rm V}$ $\sim$ $50-100$ mag \citep[e.g.,][]{wil83}. The mean gas number density of L1688 is approximately a few $\times 10^{3} \, \rm{cm ^{-3}}$. 
\begin{figure}[ht!]
\begin{center}
	\includegraphics[height=7cm,width=8.5cm,trim=5mm -5mm 5mm -5mm, clip=true]{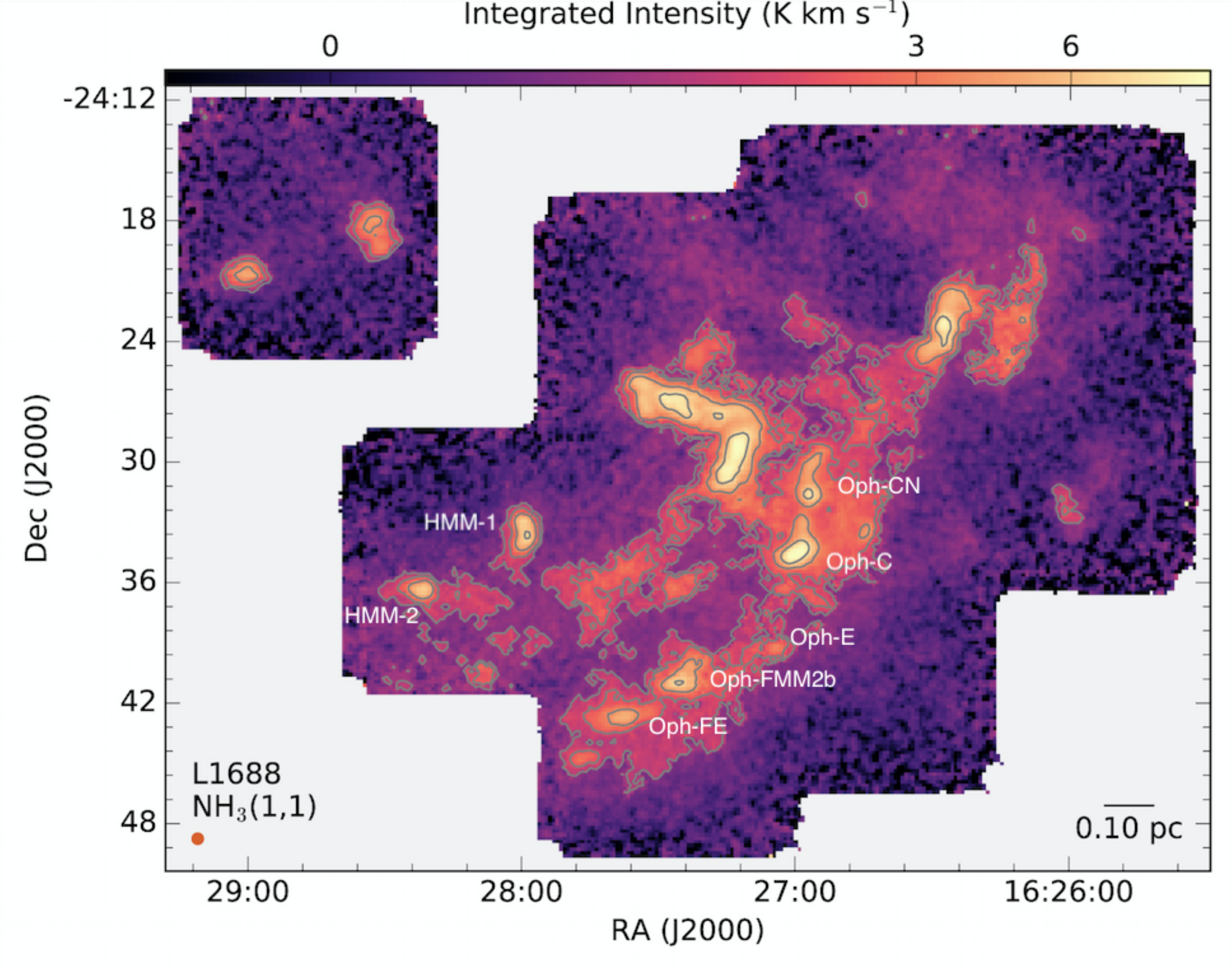}
    \caption{Integrated intensity map of the NH$_{3}\,(1, 1)$ line for the L1688 region taken from \cite{fri17}. The beam size and scale bar are shown in the bottom left and right corners respectively. The cores studied in this paper are indicated by name. }
    \label{l1688_mom0}
\end{center}
	\end{figure}
    Submillimeter continuum emission from dust shows that the star formation efficiency of the dense gas cores is $ \approx 14 \% $ \citep{jaz08}.
Figure \ref{l1688_mom0}  \citep[modified from Figure 6 in][]{fri17} shows the integrated intensity map of the NH$_{3}(1,1)$ line for the L1688 region in Ophiuchus, along with the marked cores that are studied in this paper. The map includes four prominent isolated starless cores (including H-MM1 and H-MM2) lying on the outskirts of the cloud, plus more than a dozen local line width minima in the main cloud (mainly in the south-eastern part in regions called Oph-C, E, and F). Many of these minima correspond to roundish starless cores that can be identified on the SCUBA-2 850 micron dust continuum map.
The cores indicated as Oph-C, Oph-CN, Oph-E, and Oph-FE  correspond to source names C-MM3, C-MM11, E-MM2d, and  F-MM11 respectively, as mentioned in \cite{pat15}. Furthermore, Oph-C, Oph-E, and HMM-1 are classified as starless, while the other cores are known to have a protostar (see Table 1 in \cite{pat15} for more core properties). 

\begin{figure}[ht]
\centerline{\includegraphics[height=6cm,trim=0mm 0mm 0mm 0mm, clip=true]{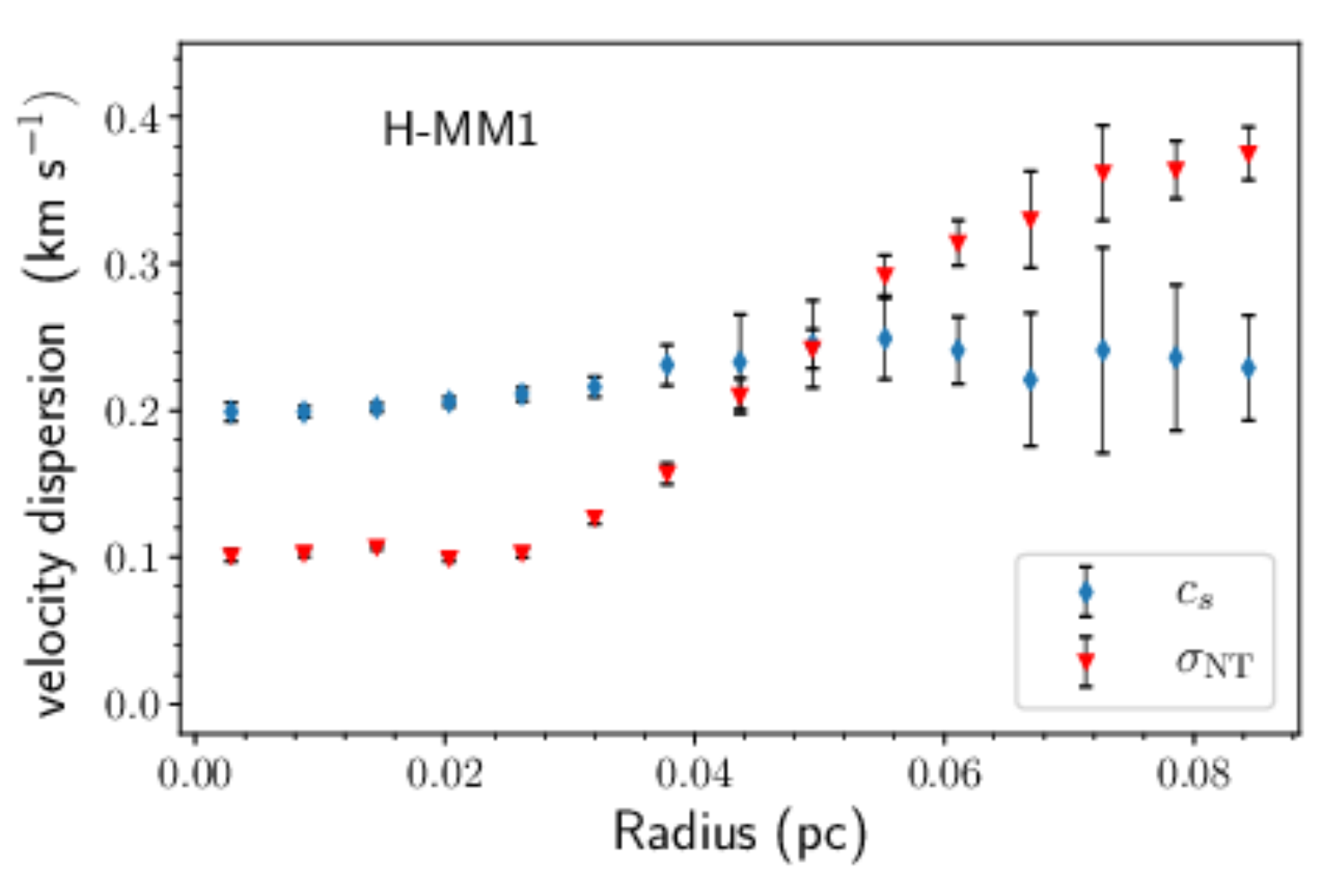}}
\caption{Velocity dispersion profiles of the H-MM1 core calculated using the NH$_{3}\,(1,1)$ and NH$_3\,(2,2)$ spectral line cubes from the GAS survey of the L1688 region of Ophiuchus \citep{fri17}. Here $c_{s}$ and $\sigma_{\rm{NT}}$ are calculated from annularly averaged spectra, by first aligning the spectra in velocity with the help of the $v_{\rm{LSR}}$ map. The transonic radius (where $c_{s}= \sigma_{\rm{NT}} $) of H-MM1 can be identified as $r_{0} \simeq 86 '' $ (0.05 pc at the $120$ pc distance of Ophiuchus). \label{veldisp}}
\end{figure}

\begin{figure*}[ht!]
\begin{center}
	\includegraphics[height=10cm,width=7.2in,trim=0mm 0mm 0mm 00mm, clip=true]{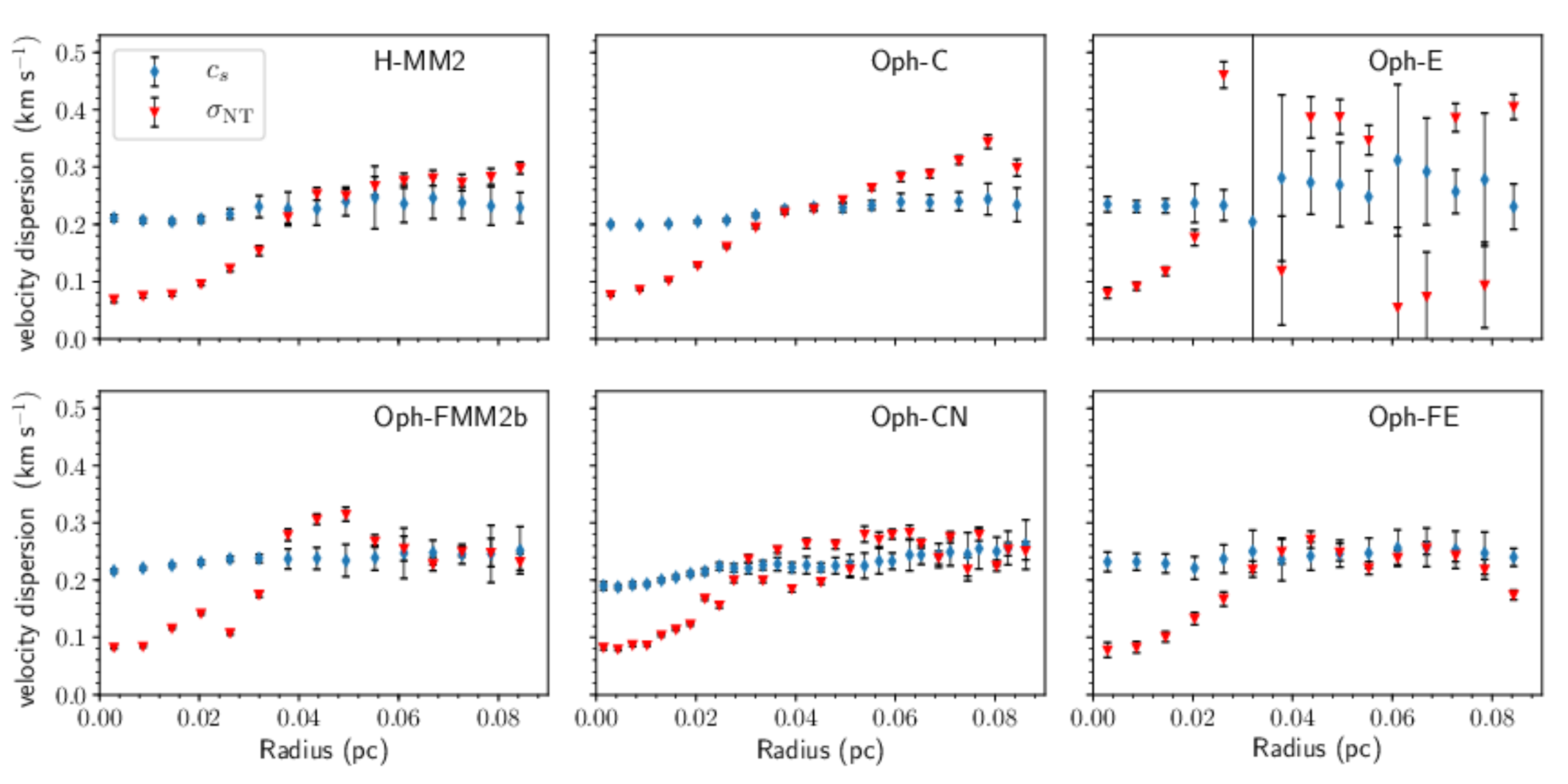}
    \caption{ Velocity dispersion profiles of some selected cores in Ophiuchus calculated using the NH$_{3}\,(1,1)$ and NH$_3\,(2,2)$ spectral line cubes from the GAS survey of the L1688 region of Ophiuchus \citep{fri17}. Here $c_{s}$ and $\sigma_{\rm{NT}}$ are calculated from annularly averaged spectra, by first aligning the spectra in velocity with the help of the $v_{\rm{LSR}}$ map.}
    \label{vel_dispersion_all}
\end{center}
	\end{figure*}

\begin{table*}
	\centering
	\caption{Core properties derived from the submillimeter continuum observations and NH$_3$ lines observations.}
    
	\label{table}
	\begin{tabular}{lcccccccc} 
		\hline
		\hline
		Core  & RA $16^{\rm h}$ & Dec $-24 ^{\rm h}$ & $r_{\rm c} \,$ & $(\sigma_{\rm NT})_{\rm c} \, $& $n_{0}\,$ & $r_{\rm 0} \, $& $p/2$  & Mass  \\ name & (J2000) & (J2000) & $({10 ^{-2 }\,\rm pc})$ & $\rm{(km\,s^{-1})}$ &  $( \,10^{5 } \,\rm{cm^{-3})}$ & $ \rm{(10 ^{-2}\, pc)}$ & & (M$_{\odot})$ \\
		
		\hline
		H-MM1 & 27:58.56 & 33:39 & 5.0 & $0.25$ & 8.0 $\pm$ 3.0 &   1.0 $\pm$ 0.4 & 1.3 $\pm$ 0.2  & 1.7 $\pm$ 0.8    \\
		H-MM2 & 27:28.21 &36:27 &  3.9 & $0.23$ & 9.0 $\pm$ 2.0  & 1.0 $\pm$ 0.2 & 1.4 $\pm$ 0.2 & 0.9 $\pm$ 0.3\\
		Oph-C & 26:59.40 &34:25& 4.5 & $0.23$  & 7.0 $\pm$ 3.0 & 3.0 $\pm$  1.0  & 1.4 $\pm$ 0.5& 5.1 $\pm$ 2.1\\
		Oph-E & 27:05.80 & 39:19 & 2.2  & $0.24$ & 8.0 $\pm$ 3.0  & 0.8 $\pm$ 0.3 & 0.9 $\pm$ 0.1& 0.6 $\pm$ 0.2 \\
		Oph-FMM2b & 27:25.10 & 41:00 &  3.5  & $0.24$ & 18.0 $\pm$ 8.0 & 0.5 $\pm$ 0.2& 0.8 $\pm$ 0.1 & 1.5 $\pm$ 0.7 \\
		Oph-CN & 26:57.10 & 31:47 &  2.9 & $0.22$ & 4.0 $\pm$ 1.0  & 1.0 $\pm$ 0.5 & 0.8 $\pm$ 0.1 & 0.7 $\pm$ 0.2 \\
		Oph-FE & 27:45.80 & 44:40 & 3.6  & $0.23$ & 2.0 $\pm$ 0.9 & 1.0 $\pm$ 0.4 & 0.9 $\pm$ 0.2 & 0.5 $\pm$ 0.2\\
		\hline
	\end{tabular}
    \tablecomments{The Plummer fit parameters are $n_{0}$, $r_{0} $, and $p/2$, $r_{\rm c}$ is the transonic radius and $(\sigma_{\rm NT})_{\rm c}$ is the velocity dispersion at $r_{\rm c}$. The final column gives an estimate of the mass of each core.}
\end{table*}    
\begin{figure*}

\gridline{\fig{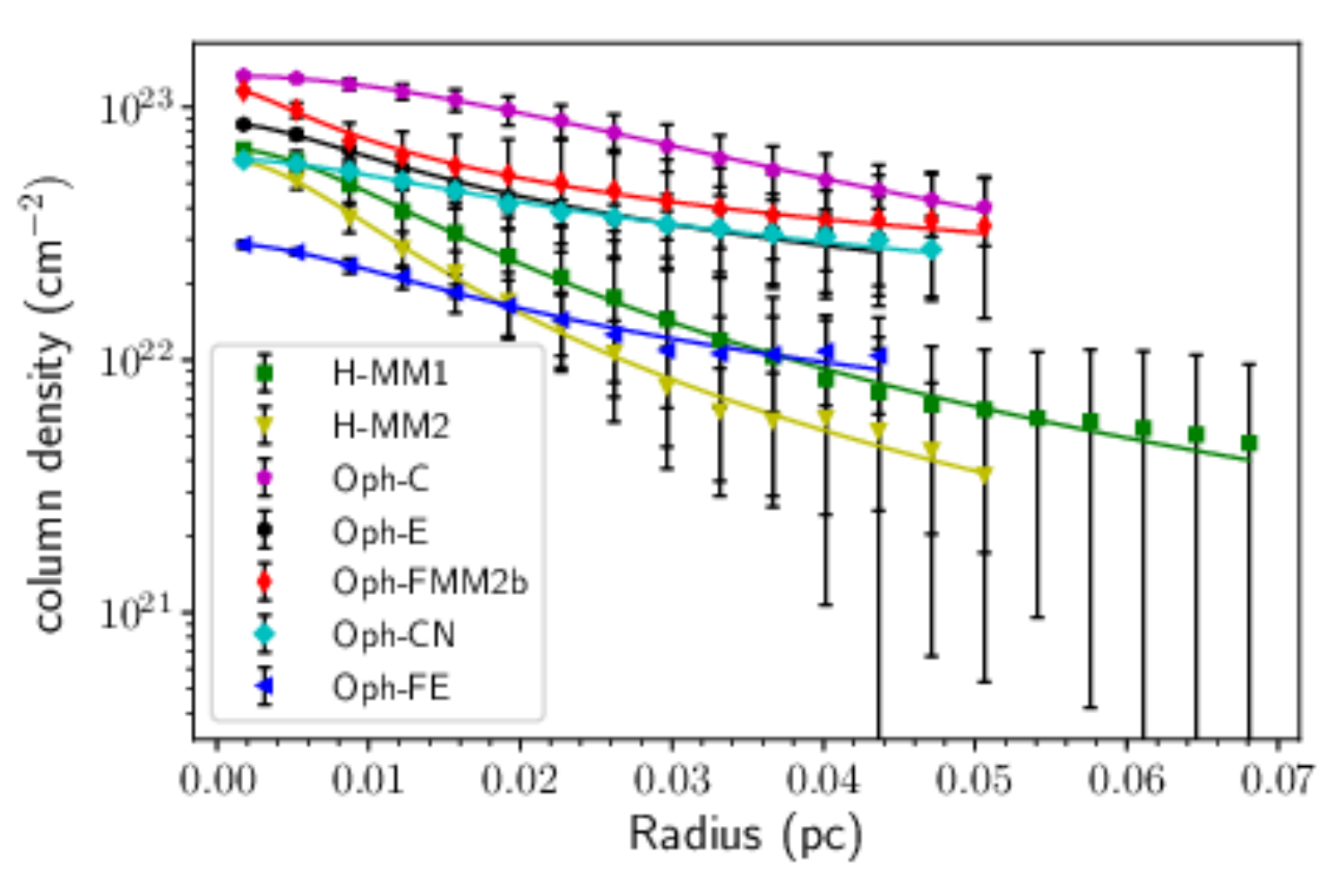}{.51\textwidth}{(a)}
          \fig{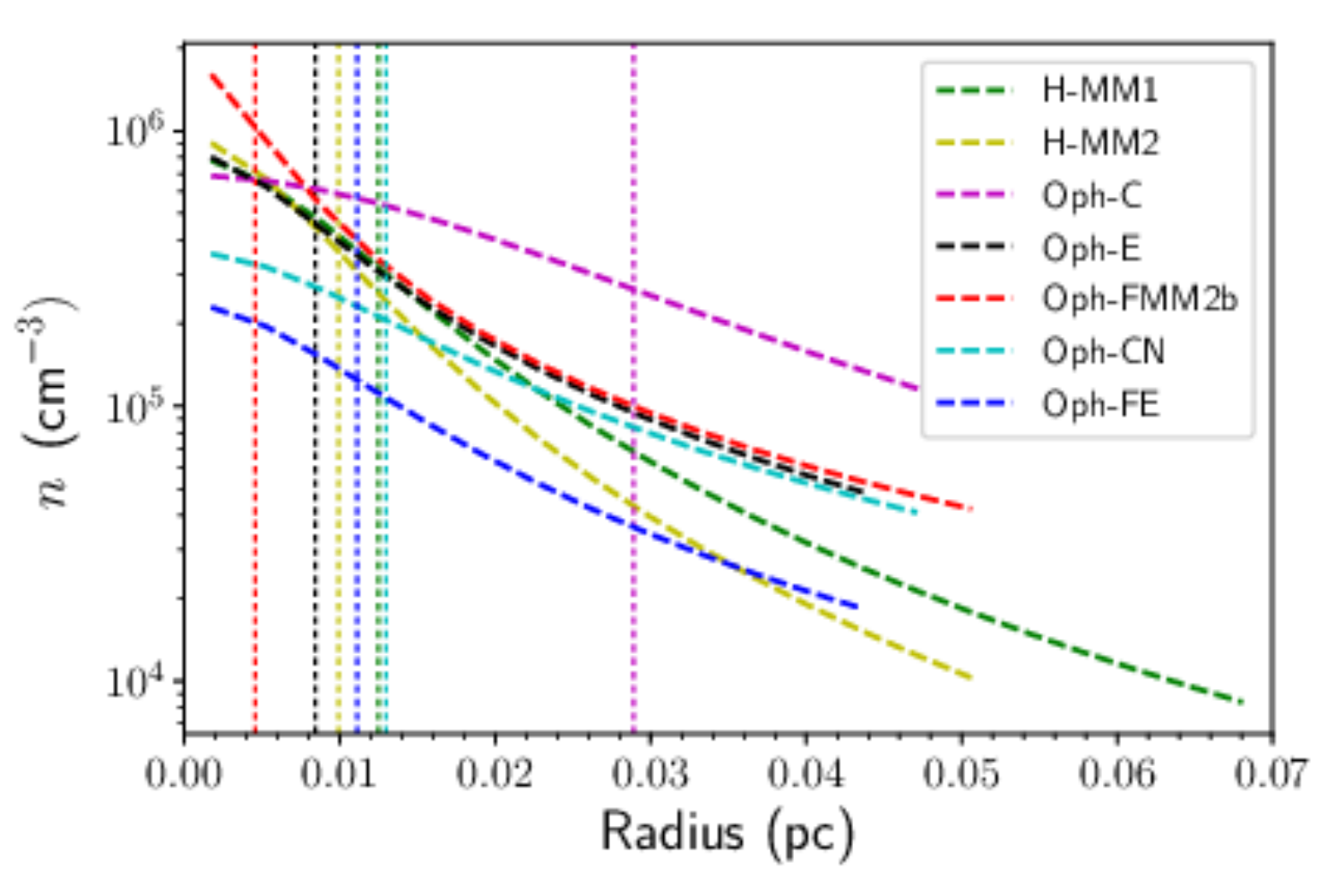}{0.51\textwidth}{(b)}}

\caption{Left: Submillimeter intensities as functions of radial distance from the center of cores in Ophiuchus. The colored markers with error bars indicate averages over concentric annuli and their standard deviations. These are obtained from SCUBA-2 maps at 850 $\mu$m published by \cite{pat15}. The solid curves are fits to the data using the Plummer model. Right: The density profiles as functions of radial distance from the center of cores in L1688. Here we plot $n(r)$ using the fit parameters (see Table \ref{table}). The vertical dotted lines mark the extent of the central flat region $r_{0}$.} 
\label{column density_set}

\end{figure*}     

\subsection{Velocity Dispersion }\label{sec:results}
The radial distributions of the velocity dispersions and the kinetic
temperatures were calculated from aligned and averaged NH$_{3}(1,1)$ and
NH$_{3}(2,2)$ spectra. The averages were calculated in concentric annuli,
weighting the spectra by the inverse of the rms noise (for example see Figure \ref{averagedsepec}, which demonstrates annular averaging of spectra for the H-MM1 core). Before
the averaging, the spectra were aligned in velocity using LSR
velocity maps produced by the reduction and analysis pipeline for the
Greenbank K-band Focal Plane Array Receiver \citep{mas11,fri17}\footnote{The data are available through https://dataverse.harvard.edu/dataverse/.}. 

The stacked NH$_{3}(1,1)$ and $(2,2)$
spectra were analyzed using the standard method described by \cite{ho83} and, recently, by \cite{fri17}. In this method, the
velocity, line width, the total optical depth, and the excitation
temperature of the $(1,1)$ inversion line are determined simultaneously
by fitting a Gaussian function to the 18 hyperfine components.  The
assumption is that individual hyperfine components have equal
excitation temperatures, T$_{\rm {ex}}$, beam-filling factors, and
line widths.

The column density of molecules in the $(J=1,K=1)$ level
depends on T$_{\rm {ex}}$ and is proportional to the product of the line width
and the total optical depth.  The $(2,2)$ inversion line is usually optically
thin, and the column density of the molecules in the $(J=2,K=2)$ level is estimated using the
integrated intensity of the $(2,2)$ inversion line. It is assumed that the $(2,2)$ inversion line 
has the same dependence on T$_{\rm {ex}}$ and has the same relation to the line width
as the $(1,1)$ line. The ratio of the column densities
of the $J,K=(1,1)$ and $(2,2)$ levels defines the rotation temperature,
T$_{\rm rot}$.  The kinetic temperature T$_{\rm kin}$ was estimated using the three
level approximation, including levels $J,K=(1,1)$, $(2,2)$, and $(2,1),$ as
described by \cite{wal83} and \cite{dan88}.

The nonthermal velocity dispersion in an averaged spectral line was
calculated by subtracting in quadrature the thermal velocity
dispersion of ammonia molecules from the total velocity
dispersion. The errors of the thermal and nonthermal velocity
dispersions were calculated by propagating the uncertainties of the
variables derived from the averaged spectra. Here it is assumed that the error in T$_{\rm kin}$ does not correlate with that of the line width. The
dominant uncertainties in the T$_{\rm kin}$ estimate are related to the
optical thickness of the $(1,1)$ line (depending on the relative
intensities of the hyperfine components) and the integrated intensity
of the $(2,2)$ line. The relative error of the line width is usually very
small (a few percent) and has a minor effect on the uncertainty in
T$_{\rm kin}$.

Figure \ref{veldisp} shows the radially averaged isothermal sound speed $c_{\rm s}$ and nonthermal velocity dispersion $\sigma_{\rm NT}$ in HMM-1. Here $c_{\rm s} = \sqrt{k T / \mu \, m_{\rm H}} $ where $T$ is the kinetic temperature, $m_{\rm H}$ is the mass of a hydrogen atom and $\mu  = 2.33$ is the mean molecular mass. Furthermore, $\sigma_{\rm NT} = \Delta v_{\rm NT} /\sqrt{2 \ln 2}$, and $\Delta v_{\rm NT}$ is the nonthermal contribution to the NH$_3$ line width. There is a clear transition point at radius $\approx 86 ^{\prime\prime}$, where $c_{\rm s} = \sigma_{\rm NT}$. We identify this radius as the transonic radius, $r_{\rm c}$, and consider it to be the core boundary. The nonthermal velocity dispersion is $\approx 0.5 c_{\rm s}$ inside the core, and it increases steeply to $\approx 2 c_{\rm s}$ across the core boundary.  We use the same prescription to map the thermal and nonthermal velocity dispersion of six other selected cores in L1688. Figure \ref{vel_dispersion_all} shows the annularly averaged thermal and nonthermal velocity dispersions of all the other selected cores in L1688. We have selected only those cores that have a distinct delineation between thermal/nonthermal line-widths ($c_{\rm s} = \sigma_{\rm NT}$) at a transonic radius $r_{\rm c}$ with the nonthermal dispersion becoming subthermal towards the center of the core. Outside the transonic radius for some cores (for example Oph-CN and Oph-Fe) the nonthermal dispersion is comparable to the sound speed. 

In Table \ref{table} we give the measure of the transonic radius $r_{\rm c}$ and the corresponding velocity dispersion $(\sigma_{\rm NT})_{\rm c}$ at $r_{\rm c}$. The values of $r_{\rm c}$ and $(\sigma_{\rm NT})_{\rm c}$ are obtained by interpolating the thermal and nonthermal data points and finding their intersection. 

\subsection{Column Density and Density Model}
Figure \ref{column density_set} shows the circularly averaged 850 $\mu$m intensity profiles of seven cores in L1688 derived from SCUBA-2 maps (see Figure 1 in \cite{pat15}).  In order to characterize each observed column density profile, we adopt an idealized Plummer model of a spherical core \citep{arz11} with radial density 

\begin{equation}\label{plummer}
\rho(r) = \frac{\rho_{0}}{\left[ 1 + (r/r_{0})^{2}\right]^{p/2}},  
\end{equation}
where the parameter $r_{0}$ is the characteristic radius of the flat inner region of the density profile,  $\rho_{0} = \mu m_{\rm H} n_{0} $ is the density at the center of the core and $p $ is the power-law index. The column density profile for such a sphere of radius $r$ can be modeled as
\begin{equation}
\Sigma_{p}(r) = A_{p} \frac{\rho_{0}r_{0}}{\left[ 1+ (r/r_{0})^2\right]^{\frac{p-1}{2}}},
\end{equation}
where $\Sigma = \mu m_{\rm H} N_{\rm H_{2}}$ is the observed column density, $N_{\rm H_{2}}$ is the number column density, and
\begin{equation}
A_{p} =  \int^{\infty}_{-\infty} \frac{du}{(1+u^{2})^{p/2}}
\end{equation}
is a constant. We fit the model profile to the SCUBA-2 $850\, \mu \rm m $ data after they are averaged over concentric circular annuli. For fitting the model to the observational data, $r_{0}$, $n_{0}$ (number density at the center), and $p$ are treated as free parameters. The left panel in Figure \ref{column density_set} is the Plummer fit to the averaged submillimeter intensities of the concentric annuli of selected cores (with clearest delineation between thermal/nonthermal motions) in L1688 region in Ophiuchus. The results from the fit are summarized in Table \ref{table}. On the right panel of Figure \ref{column density_set} we plot the density profile of all the cores (using Equation (\ref {plummer})). For most of the cores there is a noticeable central flat region of nearly constant density and then a gradual power-law decrease radially outward. The index $p/2$ is different for each model and varies in the range $0.78 < p/2 \leq 1.38$. An estimate of the mass of each core is also given in Table \ref{table}. The mass is calculated by integrating the spherical density profile up to the core radius (transonic radius). We run several iterations where the density fit parameters are drawn randomly from respective normal Gaussian distributions with a standard deviation equal to the error range of each parameter. Additionally, for each core we assume a spread of $10 \%$ for the transonic radius $r_{\rm c}$ to incorporate the uncertainty (which on average is $\approx 10 \%$) in the thermal and nonthermal line widths. The obtained mass distribution is skewed. The process is repeated 100 times and the uncertainty is calculated from the mean of $S/\sqrt{(2 \ln 2)}$, where $S$ is the semi-interquartile range, for each distribution. 

\section{Model}
The CFS model assumes that magnetic field lines are effectively frozen-in to the gas, i.e., the contraction time is shorter then the time scale associated with flux loss \citep{fie93}. The field lines are pinched towards the center of the core due to gravitational contraction. Furthermore, Alfv{\'e}nic fluctuations are assumed to dominate the nonthermal component of the velocity dispersion. In this section we discuss the details of the theory and provide justifications. We apply it to the seven selected cores to predict their magnetic field strength profile, the mean magnetic field fluctuation $\delta B $ and the mass-to-flux ratio profile.

\subsection{Core Field Structure}
Our first assumption is that the field strength follows a power-law approximation due to flux freezing. The magnetic field $B(r)$ within the core radius $r_{\rm c}$ can be written in terms of the observed values as 
\begin{equation}\label{magneticfieldmodel}
B(r)/B(r_c) = [\rho(r)/\rho(r_c)]^\kappa,
\end{equation}
where $1/2 \le \kappa \le 2/3$ \citep{cru12} is a power-law index. Here $B(r_c)$ is the field strength at the transonic radius. The gas density approaches a near uniform value outside $ r_{\rm c}$. Thus, we do not extend the power-law approximation beyond the core radius. We assume that the core is truncated by an external medium as for a Bonnor-Ebert sphere.
Equation (\ref{magneticfieldmodel}) approximates various relations obtained from theoretical and numerical models of magnetic cores. \cite{mes66} showed that $\kappa = 2/3$ in the limit of weak magnetic field and spherical isotropic contraction (which can occur if thermal support nearly balances gravity). Theoretically, $B \propto \rho^{2/3} $  relates the mean field and the mean density within a given radius. In Equation (\ref{magneticfieldmodel}) we generalize that idea with the approximation that $B \propto \rho^{2/3} $ can be applied to obtain the local magnetic field $B(r)$ using the local density $\rho(r)$. This approximation has an associated uncertainty of a factor $\lesssim 2$ as discussed in section \ref{dis}. 

In the limit of gravitational contraction mediated by a strong magnetic field \cite{mou76b} showed that $\kappa $ is closer to $1/2$. In the limit of very strong magnetic field (subcritical mass-to-flux ratio) models of the ambient molecular cloud \citep{fie93}, ambipolar diffusion leads to the formation of supercritical cores within which $\kappa = 1/2$. We are only applying Equation (\ref{magneticfieldmodel}) within the transonic radius, within which local self-gravity is presumed to be dominant.



\begin{figure}
\centering
 \includegraphics[height=6.1cm,trim=4mm 0mm 4mm 0mm, clip=true]{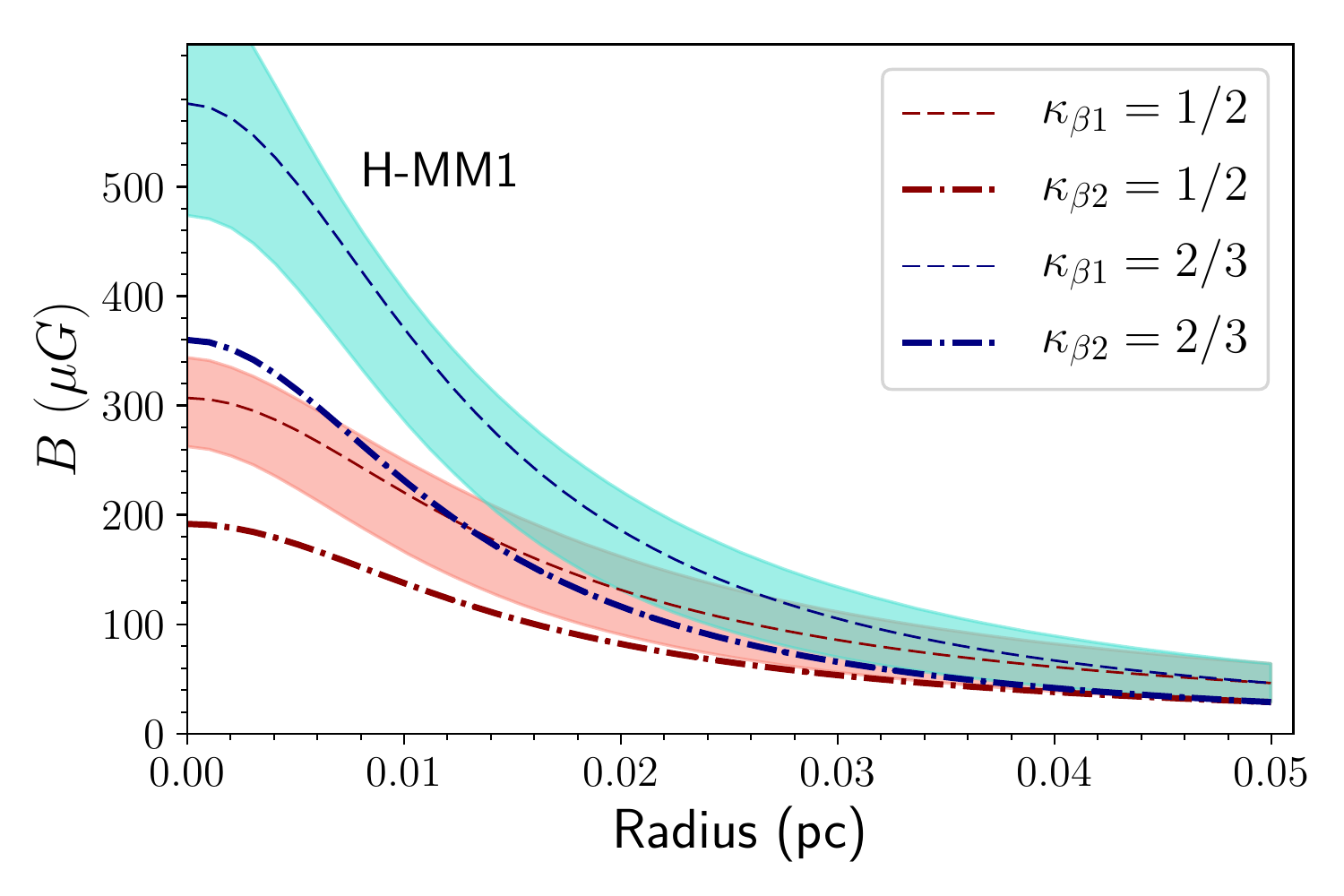}
\caption{The magnetic field profile of H-MM1 obtained from the CFS model using the observed line widths and density. The red and the blue dashed lines are the magnetic field $B$ for $\kappa = 1/2$  and $\kappa = 2/3$, respectively for the choice of $ \beta_{1}=0.5 $. The shaded region encloses the first and the third quartile of the distribution obtained using a Monte Carlo analysis. The dot-dashed red and blue lines are the magnetic field $B$ for $\kappa = 1/2$  and $\kappa = 2/3$, respectively, for $\beta_{2} =0.8$.}\label{H_MM1}
\end{figure}

\begin{figure*}[ht!]

\includegraphics[height=10cm,width=7.2in,trim=0mm 0mm 0mm 0mm, clip=False]{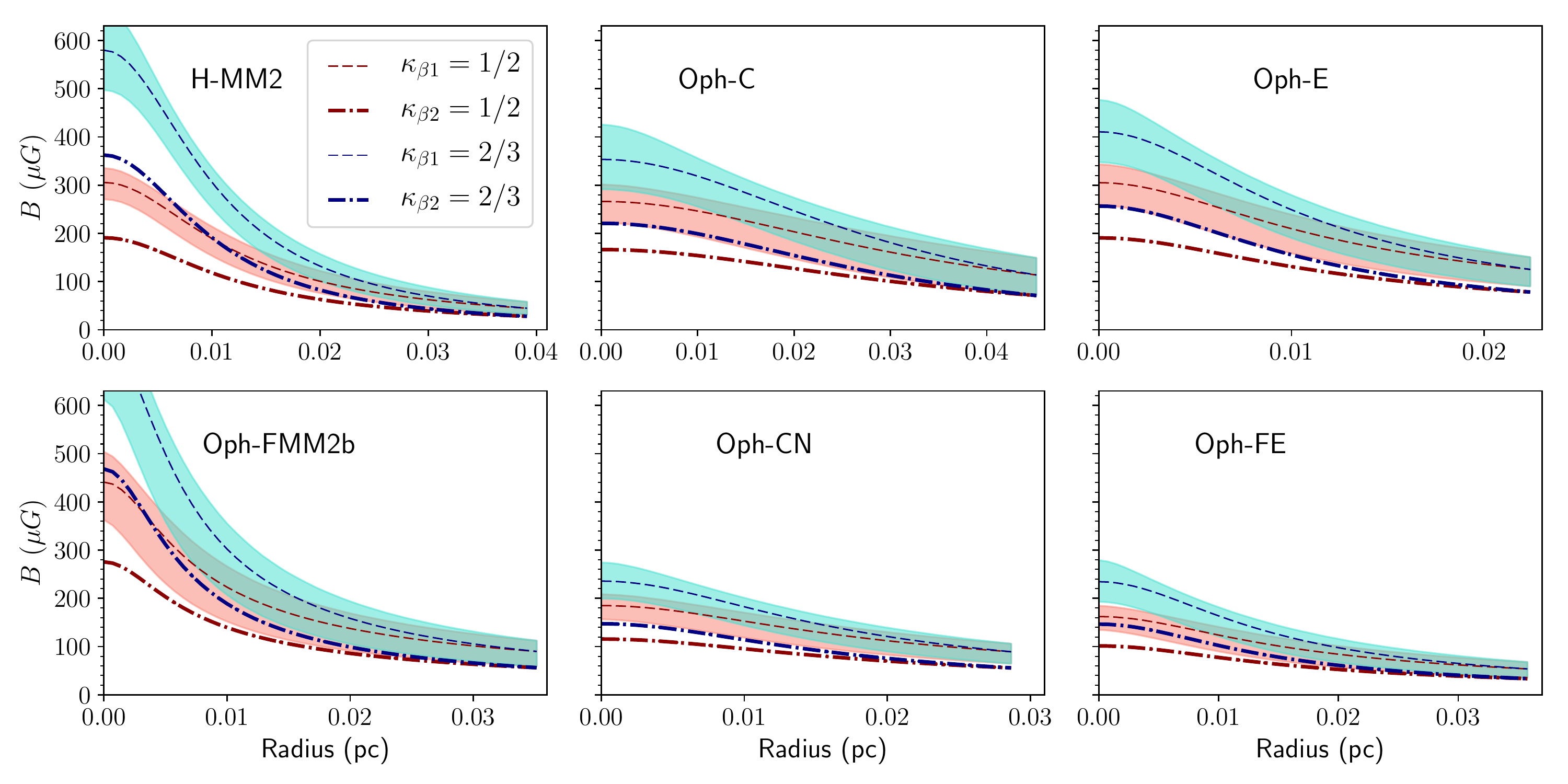}
\caption{The magnetic field profile of six different cores (names on the upper left corner) obtained from the CFS model using the observed line widths and density. The red and the blue dashed lines are the magnetic field $B$ for $\kappa = 1/2$  and $\kappa = 2/3$, respectively, for the choice of $\beta_{1} = 0.5 $. The shaded region encloses the first and the third quartile of the distribution obtained using Monte Carlo analysis. The dot-dashed red and blue lines are the magnetic field $B$ for $\kappa = 1/2$  and $\kappa = 2/3$ respectively, for $\beta_{2}=0.8 $. The magnetic field increases radially inward and the ascent is steeper for $\kappa = 2/3$. }\label{magneticfieldcores}
\end{figure*}

To model the nonthermal motions we assume Alfv{\'e}nic fluctuations. This means that we ignore possible additional sources of  the nonthermal line width, for example unresolved infall motions. The Alfv{\'e}nic fluctuations obey 
\begin{equation}\label{alfvenic}
\frac{\sigma_{\rm NT}}{v_{\rm A}} = \frac{\delta B} {B},
\end{equation}
where the Alfv{\'e}n speed is defined by $v_{\rm A} \equiv B/(\sqrt{4 \pi \rho})$. This directly leads to 
\begin{equation}\label{deltaB}
\delta B = \sigma_{\rm {NT}}\sqrt{4 \pi \rho}\,.
\end{equation}
We also use Equation (\ref{alfvenic}) to get
\begin{equation}
B(r_{\rm c}) =\frac{(\sigma_{\rm NT})_{{\rm c}}}{\beta}   \sqrt{ 4 \pi \rho(r_{\rm c}) }
\end{equation}
for use in Equation (\ref{magneticfieldmodel}) by estimating a value of relative field fluctuation $\beta \equiv \delta B /B$ at $r = r_{\rm c}$.

\cite{kud03} showed in a simulation with turbulent driving that $\beta$ is restricted to $\lesssim 1$ as highly nonlinear Alfv{\'e}nic waves quickly steepen and drain energy to shocks and acoustic motions, and that their model cloud evolved to a state in which $\sigma_{\rm NT} \approx 0.5 v_{\rm A}$. They found that for a range of different amplitudes of turbulent driving, the value of $\beta$ saturates at a maximum value in the range of $0.5$ to $0.8$. Based on these results  we pick a range $ 0.5 \le \beta \le 0.8$ at the inner boundary ($r = r_{\rm c}$) of the turbulent region. For simplicity we demonstrate only the two limiting values $\beta_{1} =0.5$ and $\beta_{2} = 0.8$ in subsection \ref{coreprop}.

The nonthermal velocity dispersion arises from transverse Alfv{\'e}nic waves. However, the observed small variation in $\sigma_{\rm NT}$ from core to core suggests that $\sigma_{\rm NT}$ is robust against the likely variation in mean field angle. This is possible because  Alfv{\'e}nic motions are  nonlinear in the outer parts of the core. Thus the Alfv{\'e}n waves will have magnetic pressure gradients (in $\delta B$) that will drive motions along the mean field direction as well. The composite nonthermal line width, accounting for
motions in all directions, is expected to be comparable to the mean
Alfv{\'e}n speed within a factor of order 2 \citep[see Figure 13 of][]{kud03}. Thus the effects of differing viewing angles are relatively small. The CFS model essentially predicts the magnetic field profiles (using Equation \ref{magneticfieldmodel}) of the dense cores in Ophiuchus.
Furthermore, it yields the variation of $\delta B /B $ and the normalized  mass-flux ratio within each core profile. We discuss some of the predicted core properties in the next subsection.

\begin{figure}
\centering
\includegraphics[height=6.0cm,trim=2mm 0mm 4mm 0mm, clip=true]{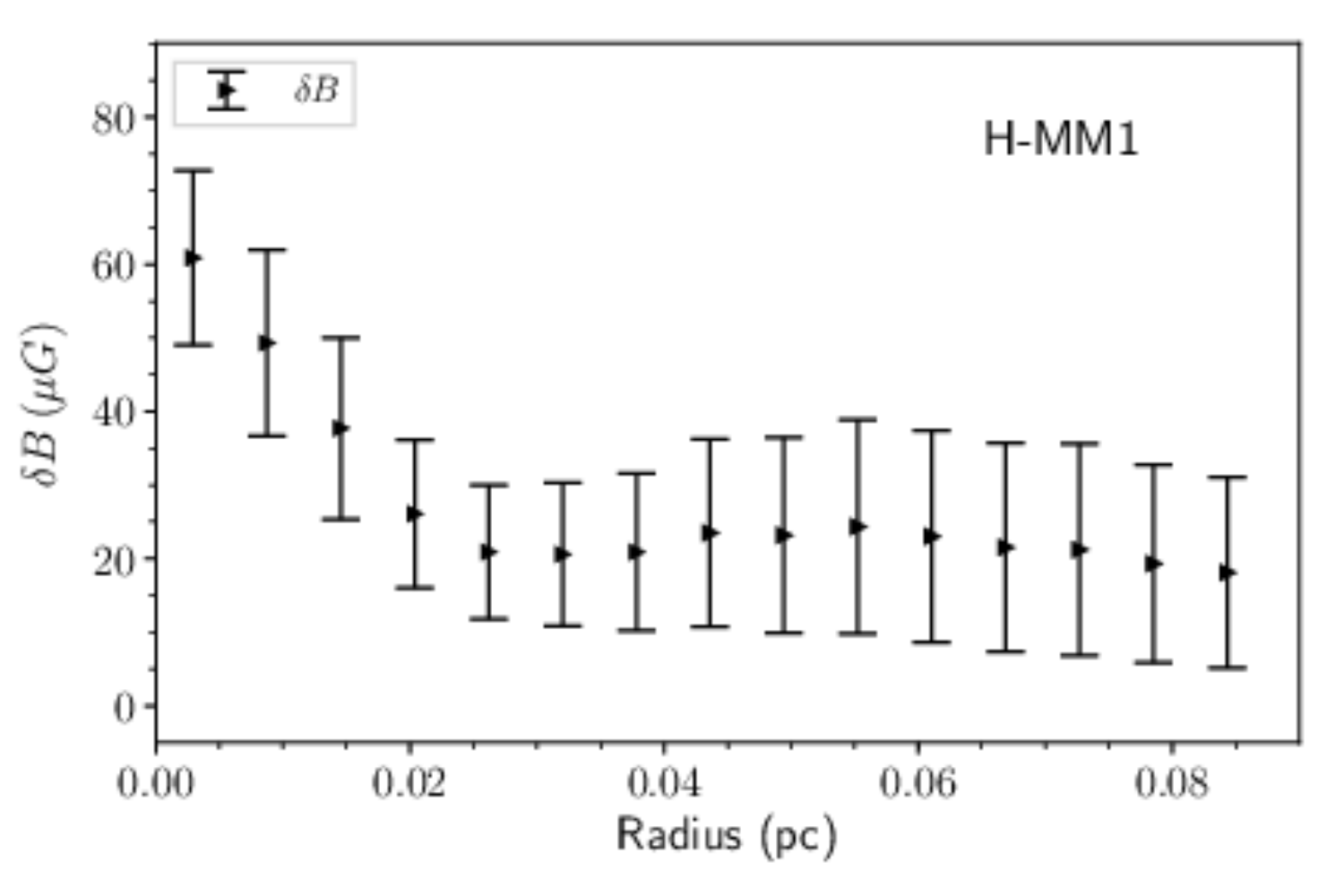}
\includegraphics[height=6.0cm,trim=2mm 0mm 4mm 0mm, clip=true]{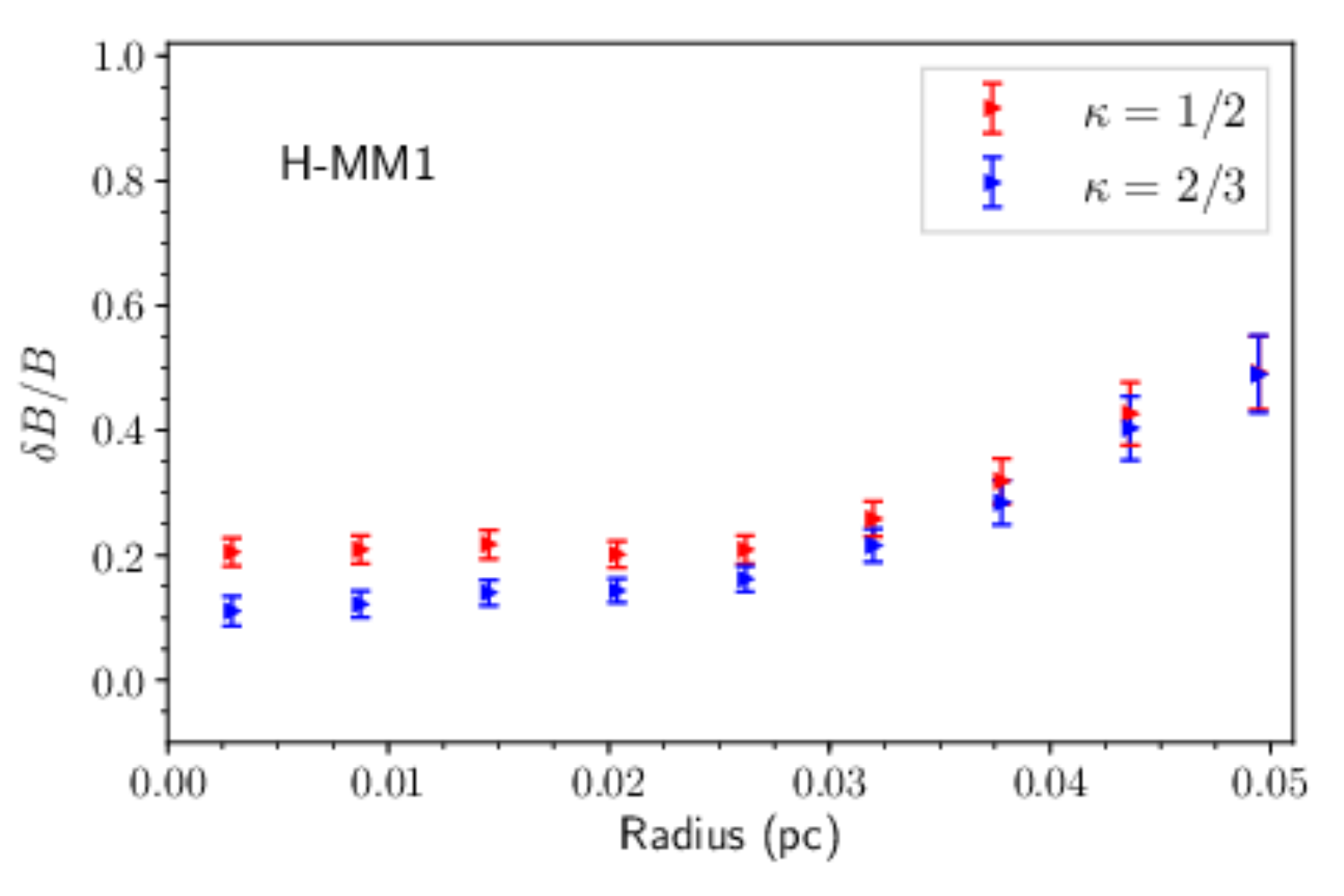}
\caption{Top: The magnetic field fluctuation $\delta B$ in H-MM1 versus radius. Bottom: The variation of $\delta B /B$ for $\kappa=1/2$ (red) and $\kappa=2/3$ (blue), respectively (assuming $\beta_1 = 0.5$). The radial profile for $\delta B /B$ is only within the transonic radius since the model (Equation \ref{magneticfieldmodel}) is applied only in that region.  The error bars in both cases are obtained using standard propagation of one sigma error and Monte Carlo analysis.}\label{changeinB}
\end{figure}




\subsection{Core Properties} \label{coreprop}
Figure \ref{H_MM1} shows the magnetic field profile of H-MM1 obtained using the CFS model. The magnetic field $B$ increases radially inward and the ascent is steeper for $\kappa = 2/3$. For example, the $B$ value at $r = 0.01 \, \rm pc$ for $\kappa = 2/3$ is $\simeq 68 \%$ greater than that for $\kappa =1/2$. Similarly, we predict the magnetic field strength profile of all the other cores using the power-law model. Figure \ref{magneticfieldcores} shows the field profile as a function of radial distance from the center. Similar to H-MM1, the field strength at a radius of $0.01 \, \rm pc$ from the center is greater for $\kappa = 2/3$ compared to $\kappa = 1/2$, with a maximum increase of $61 \%$ in H-MM2 and a minimum increase of $19 \%$ in Oph-E. Furthermore, the general increase of the field strength towards the core center can be associated with the pinching of the field lines due to flux-freezing. The power-law relation $ B \propto \rho^{\kappa}$ for $\kappa =1/2$ or $2/3$ captures different geometries. For example $ \kappa =2/3$ is consistent with a spherical core and $ \kappa =1/2$ corresponds to flattening along the magnetic field lines. 

We use a Monte Carlo analysis, where we run several iterations to evaluate the magnetic field strength using Equation (\ref{magneticfieldmodel}). The parameters (for example $r_{0}$, $\rho_{0}$, and $p/2$ ) are randomly picked from a Gaussian distribution with standard deviation equal to the error range of each parameter (see Table \ref{table}). Additionally, we assume a variation of $10 \%$ for the values of $(\sigma_{\rm NT})_{\rm c}$ and $r_{\rm  c}$ to incorporate the uncertainty (on average $\approx 10\%$) in the thermal and nonthermal line widths. The shaded region in both the plots encloses the first and the third quartile of the distribution of magnetic field strength. The dotted curve is the actual model value for $\beta_{1} = 0.5$. We repeat a similar analysis for the six other cores in Ophiuchus. There is a significant decrease in the field strength of $\approx 38 \%$ (as indicated by the dot-dashed lines) for a larger assumed value of field fluctuation (i.e., $\beta_{2} = 0.8$) at the transonic radius $r_{\rm c}$. Although there is a systematic dependence of the field strength on the choice of $\beta$, the overall shape of the magnetic field profile remains the same. 

\begin{figure*}[ht!]
\includegraphics[height=10.8cm,width=7.2in,trim=0mm 0mm 0mm 0mm, clip=False]{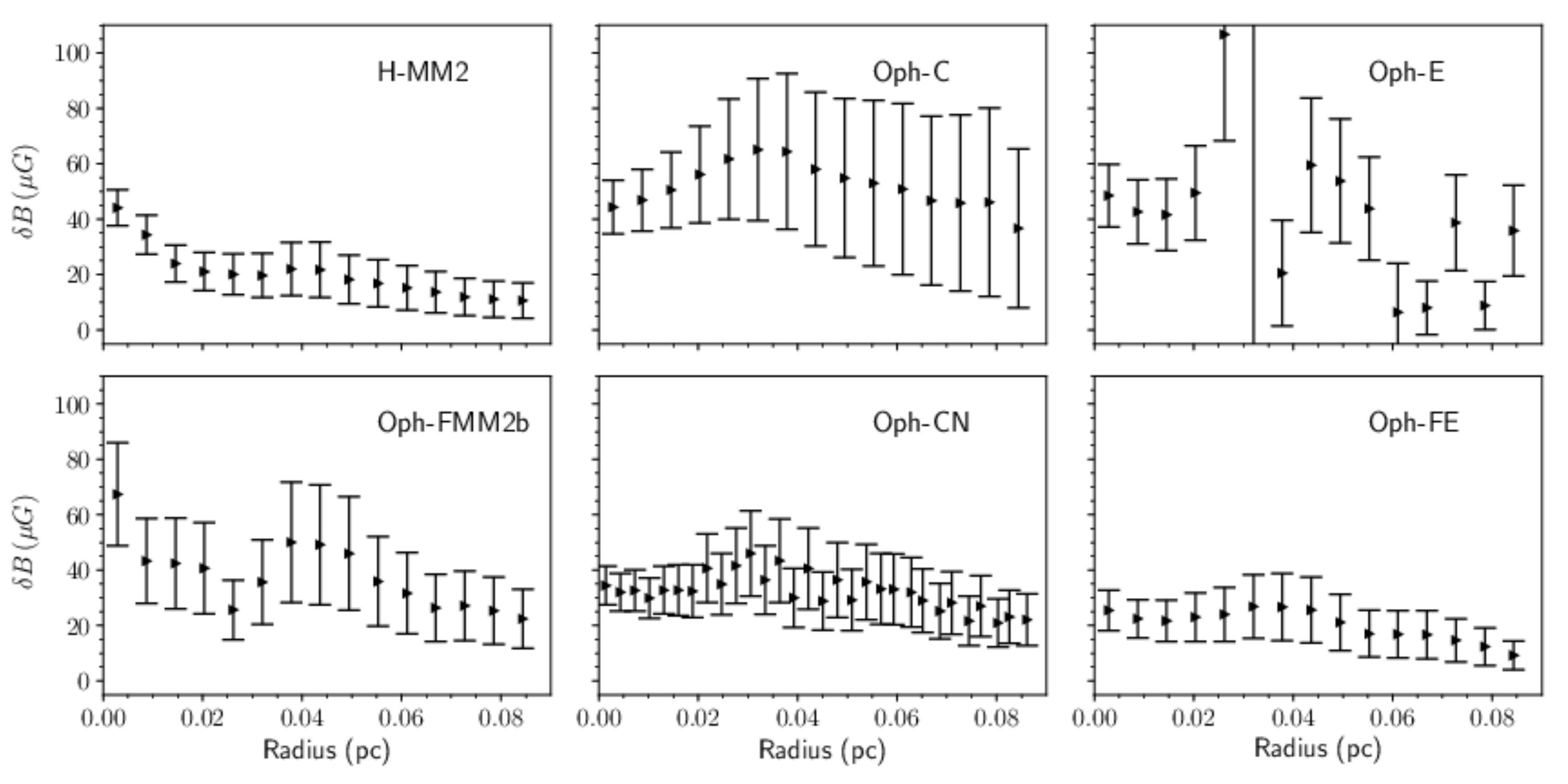}\label{delta_B_all}

\includegraphics[height=10.8cm,width=7.2in,trim=0mm 0mm 0mm 0mm, clip=False]{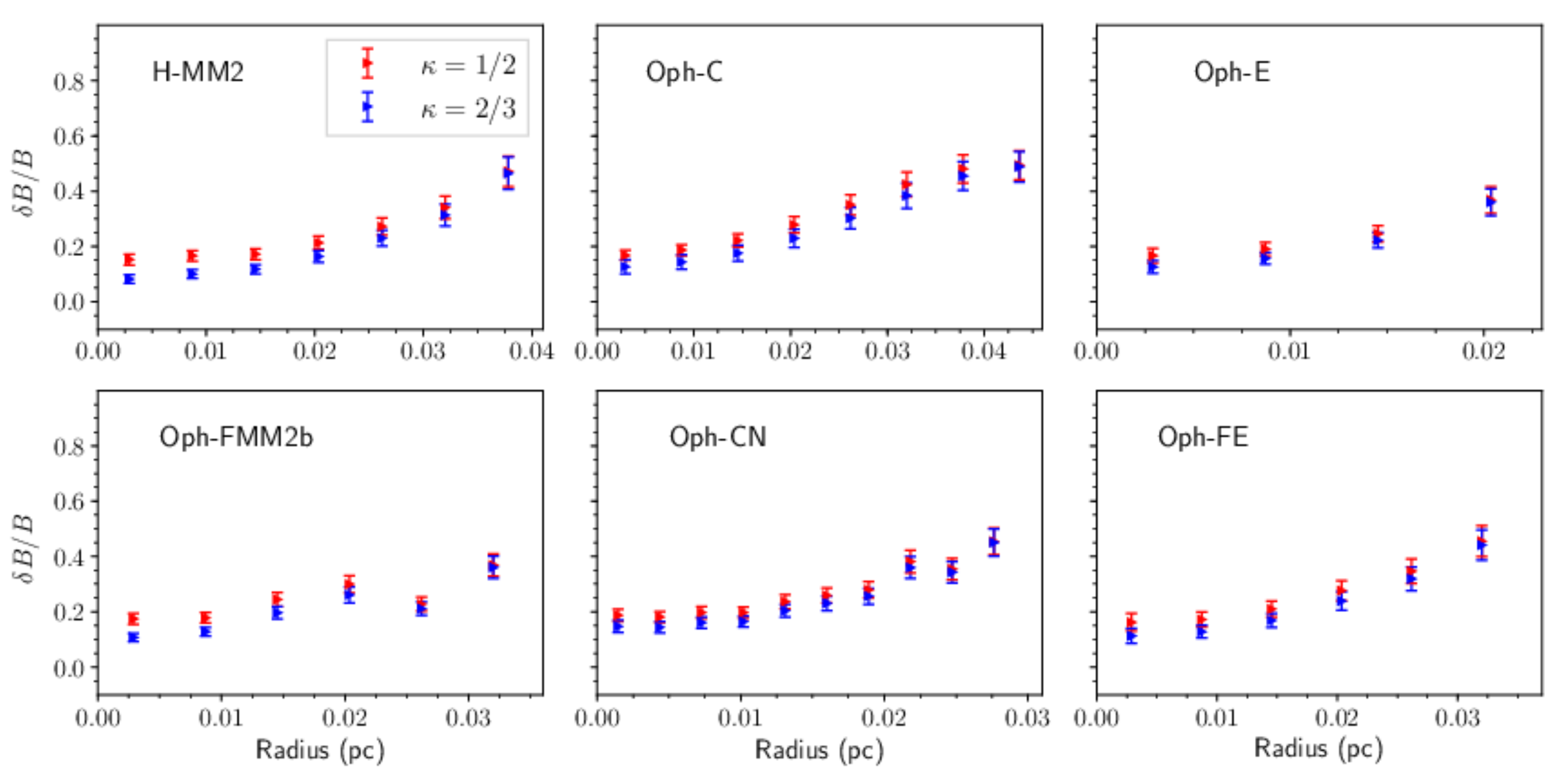} 

\caption{Top two panels: The variation of $\delta B$ across the six cores in Ophiuchus. Bottom two panels: The red and the blue lines show the variation of $\delta B /B$ for $\kappa=1/2$ and $\kappa=2/3$, respectively. The error bars in both the cases are obtained using a standard propagation of one sigma error and Monte Carlo analysis.}\label{delta_B_B_all}
\end{figure*}

Figure \ref{changeinB} shows the fluctuations of the mean magnetic field $\delta B$ and $\delta B/ B$  mapped across the H-MM1 core. These are obtained using Equation (\ref{deltaB}) and the observed nonthermal velocity dispersion data, density, and the modeled magnetic field. 
The inferred variation of $\delta B / B$ shows a trend very similar to the nonthermal velocity fluctuations. It increases outward as it approaches  the transonic radius. Inside the core $\delta B/B $ decreases to a relatively constant value of $\approx 0 .1$. The $\delta B / B$ profile essentially captures the Alfv{\'e}nic fluctuations across H-MM1. The values of $\delta B/B$ will only correspond to an observed $\delta \theta$ in polarization direction if the observed magnetic field is oriented along the plane of the sky. Figure \ref{delta_B_B_all} shows $\delta B$ and $\delta B / B$ for the other cores in Ophiuchus. They all exhibit a very similar trend as H-MM1. 

\subsection{The mass-to-flux ratio}
In this section, we estimate the normalized mass-to-flux ratio $\mu \equiv M/\Phi/(M/\Phi)_{\rm crit}$, where $(M/\Phi)_{\rm crit} = (2 \pi \sqrt {G})^{-1}$ \citep{nak78},  of the seven cores studied in this paper, assuming a spherically symmetric density profile.
The relative strength of gravity and the magnetic field is measured by the mass-to-flux ratio $M/\Phi$. For $M/\Phi > (M/\Phi)_{\rm crit}$, the cloud is supercritical and can collapse if there is sufficient external pressure. However, for $M/\Phi < (M/\Phi)_{\rm crit}$ the cloud is subcritical and the field can prevent its collapse as long as magnetic flux freezing applies. An analytic expression for $M/ \Phi$ is possible if we assume that the magnetic field lines are threading a spherical core in the plane of the sky. See Appendix \ref{appA} for the derivation.

However, for a near-flux-frozen condition, the field lines are pinched toward the central region of the dense core and resemble an hourglass morphology \citep{gir06,ste13}. We rewrite the density profile of Equation (\ref{plummer}) in normalized cylindrical coordinates $\xi \equiv x/r_{0}$ (we use $x$ as the radial coordinate) and $\zeta = z/r_{0}$ so that

\begin{figure}

\includegraphics[height=6.0cm,trim=2mm 0mm 4mm 0mm, clip=true]{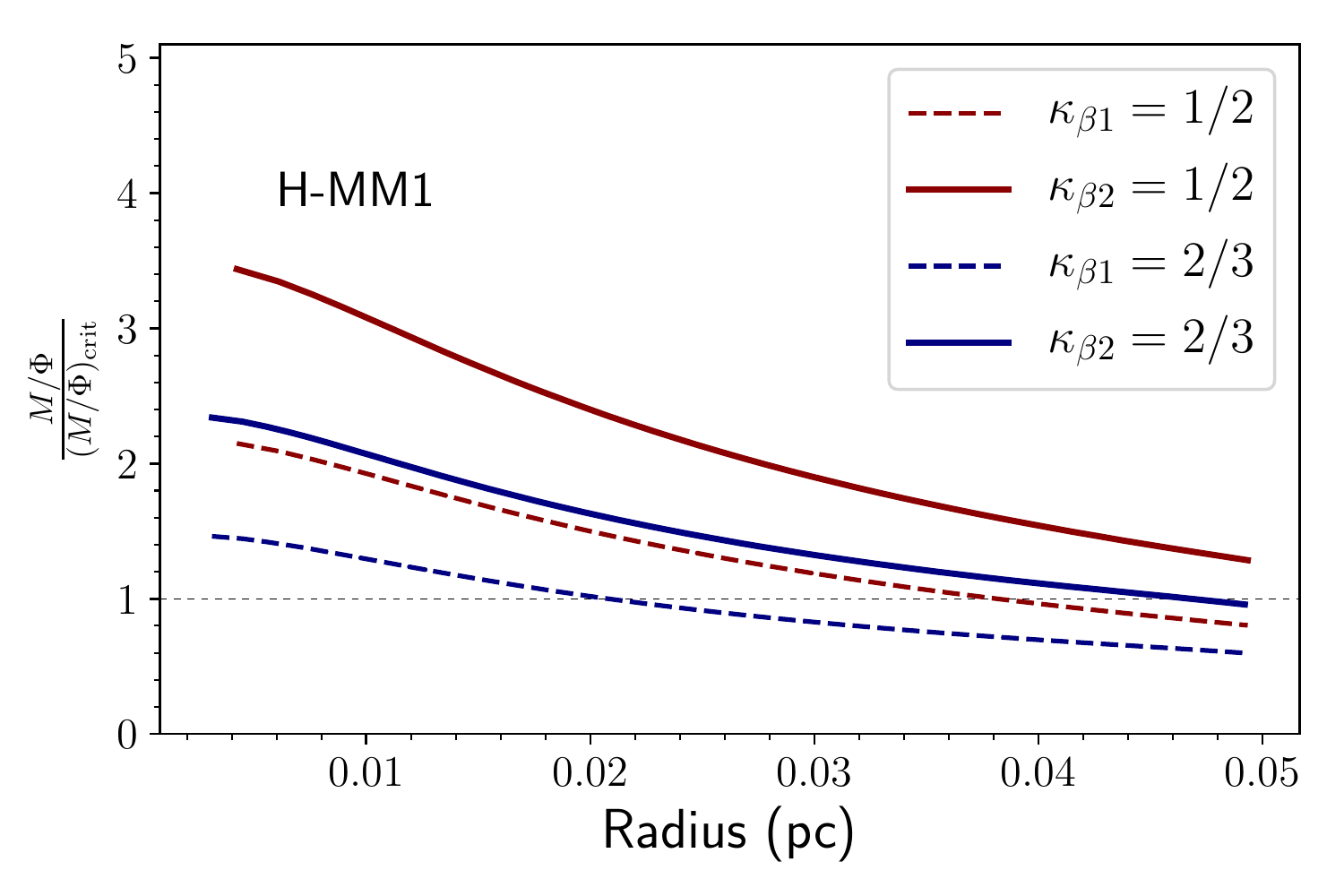}
\caption{The normalized mass-to-flux ratio $\mu \equiv M/\Phi/(M/\Phi)_{\rm crit}$ of H-MM1 as a function of radial distance from the center. The dashed and the solid lines are for $ \beta_{1}=0.5$ and $ \beta_{2}= 0.8$ respectively. The core is mostly supercritical with $\mu > 1$ (depending on the value of $\beta$) and is decreasing outward. The dotted horizontal line indicates the critical mass-to-flux ratio. }\label{mass2flux}
\end{figure}

\begin{figure*}[ht!]

\centering
\includegraphics[height=10cm,width=7.2in,trim=0mm 0mm 0mm 0mm, clip=False]{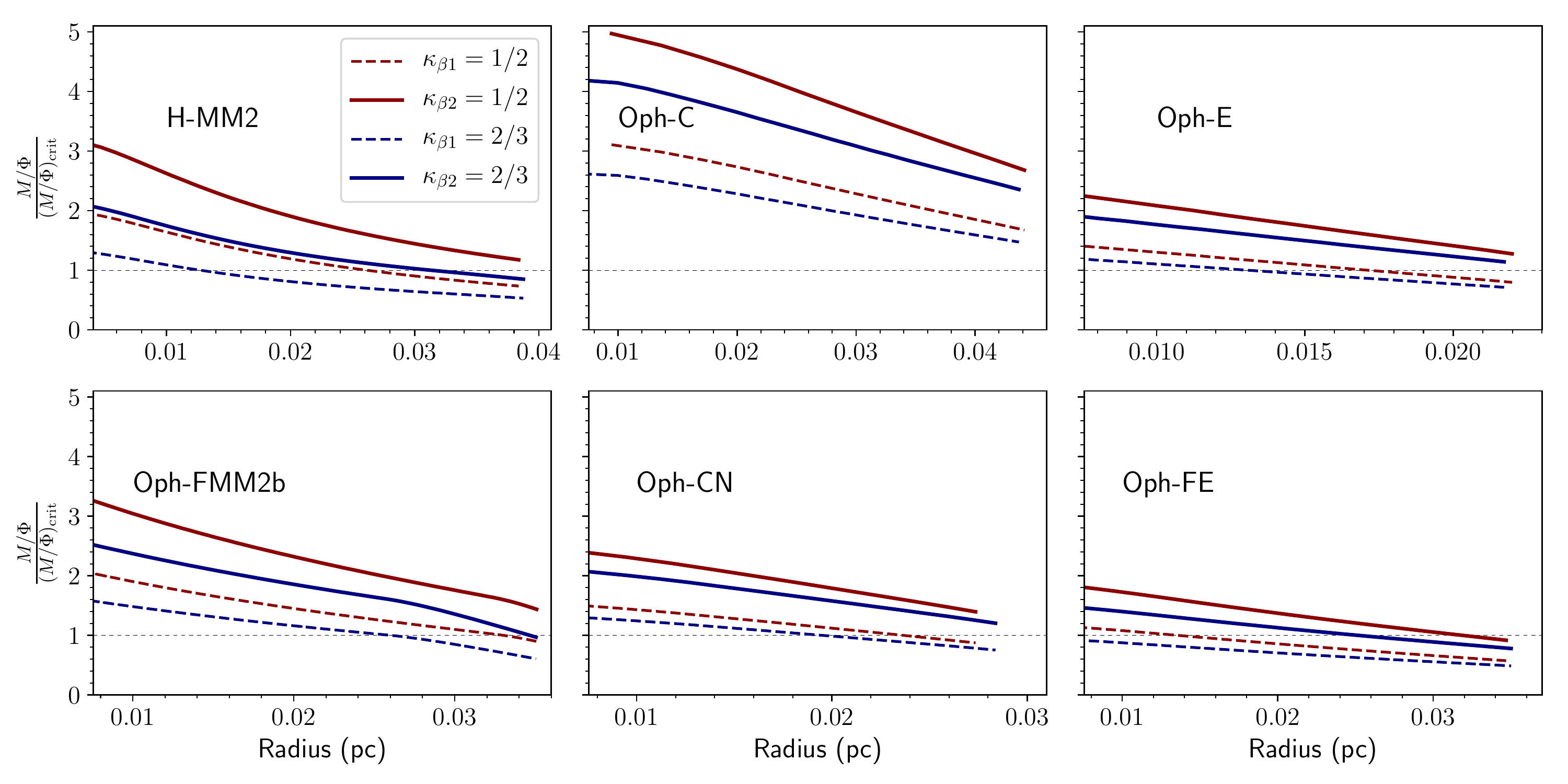}
\caption{The normalized mass-to-flux ratio $\mu \equiv M/\Phi/(M/\Phi)_{\rm crit}$ as a function of radial distance from the center for the remaining six cores in our sample. The dashed and the solid lines are for $\beta_{1} =0.5$ and $ \beta_{2} = 0.8$ respectively.  The core is mostly supercritical with $\mu > 1$ (depending on the value of $\beta$)  and is decreasing outward. The dotted horizontal line indicates the critical mass-to-flux ratio. }\label{mass2flux_error_all}
\end{figure*}

\begin{equation} \label{densityratio}
\frac{\rho(\xi,\zeta)}{\rho_{\rm c}} = \frac{\rho_{0}/\rho_{\rm c}}{\left[ 1 + \xi^{2}+\zeta^{2}\right]^{p/2}}.    
\end{equation}
Using Equation (\ref{magneticfieldmodel}) we can estimate the flux function
\begin{eqnarray}\label{flux}
\Phi (\xi, \zeta) =  2\pi r_{0}^{2} B(r_c)  \int^{\xi}_{0}\left[ \frac{\rho_{0}/\rho_{\rm c} \, \, }{\left[ 1 + \xi^{\prime 2}+\zeta^{2}\right]^{p/2}}\right]^{\kappa} \xi^{\prime} d\xi^{\prime}.
\end{eqnarray}
Here we make the approximation that at each height the flux can be estimated from the scalar magnetic field strength obtained from Equation (\ref {magneticfieldmodel}) rather than the local vertical component of $B$. This is equivalent to assuming that the field lines are not highly pinched in the observed region of the prestellar cores that are modeled here. An analytic solution to the above equation is only possible for the case where $p/2 =1$. We solve the above integral numerically and draw contours of constant magnetic flux. We note that each field line is a contour of constant enclosed flux  \citep[see][]{mou76a}.  To estimate the enclosed mass through each of the flux tubes within the core we integrate numerically. Figure \ref{mass2flux} shows the mass-to-flux ratio of H-MM1 as a function of radial distance from the center. As evident for both $\kappa = 1/2$ and $2/3$, the mass-to-flux ratio ($\mu$) is supercritical at the center and declines towards the core edge. However, the mass-to-flux ratio depends on the choice of $\beta$. See Section \ref{dis} for further discussion. Figure \ref{mass2flux} shows that for a greater value of $\beta$ the mass-to-flux estimate increases and the entire core is supercritical. Figure \ref{mass2flux_error_all} shows the profile of the mass-to-flux ratio for the six other cores studied in this paper. Most cores (namely H-MM2, Oph-E, Oph-FMM2b and Oph-CN) show a similar decline of the mass-to-flux ratio towards the transonic radius. Oph-C is supercritical all the way to the core boundary for both the $\beta$ values. Oph-FE is close to the critical limit. 

\begin{figure}

\includegraphics[height=3.5in,width=3.5in,trim=4mm 6mm 8mm 10mm, clip=true]{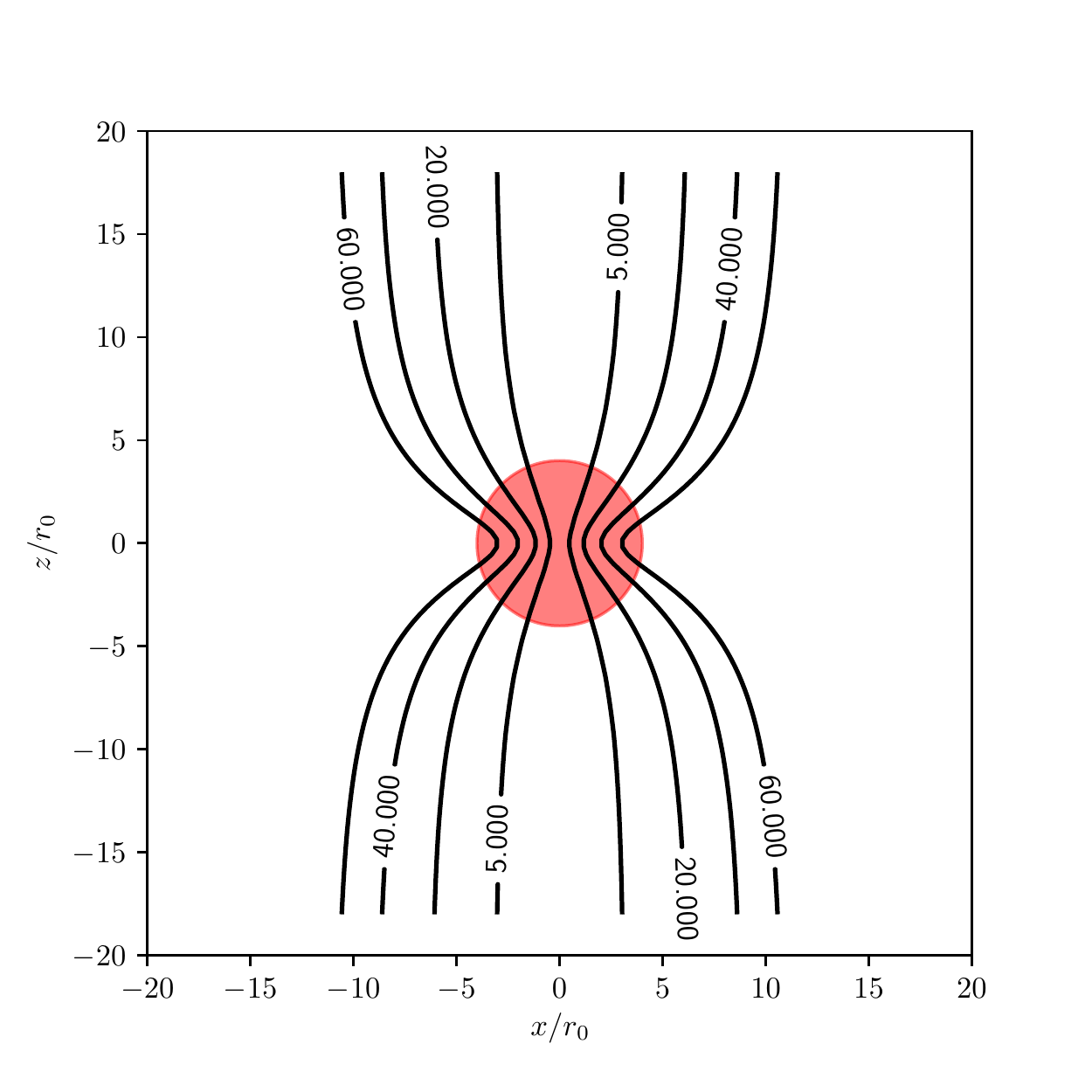}
\caption{An illustration of the magnetic flux contours in H-MM1 for the power-law model with $\kappa =2/3$. The core parameters for H-MM1 ($n_0,\,r_{0},\,r_{\rm c}$) are taken from Table \ref{table}. The peak density $\rho_{c}$ is chosen to be 300 times the background. The marked flux lines are normalized to a background value $\Phi_{0} = 2\pi r_{0}^{2} B_{\rm {u}}$ (refer to the text for details). The $x-$ and $z-$ axes are in units of $r_0 = 0.012$ pc. The circle at the center represents the H-MM1 core of radius $r_c = 0.05$ pc.}\label{H_mm1hourglass}
\end{figure}
As an example of the magnetic field lines we demonstrate the case of H-MM1, where we plot in Figure \ref{H_mm1hourglass} the flux contours for the power-law model with index $\kappa=2/3$. To represent the field lines we introduce a background field strength ($B_{\rm u}$) and background density $\rho_{\rm u}$. The flux $\Phi$ is estimated using the Equation (\ref{flux}) but with the modified density expression normalized to background density $\rho_{\rm u}$:
\begin{equation}
\frac{\rho(r)}{\rho_{\rm u}} = 1+ \frac{\rho_{0}/\rho_{\rm u}}{\left[ 1 + \xi^{2}+\zeta^{2}\right]^{p/2}}.  
\end{equation}
Here the background density $\rho_{\rm u}$ is added to the core density $\rho (r)$. The flux lines in Figure \ref{H_mm1hourglass} are normalized to $\Phi_{0} = 2\pi r_{0}^{2} B_{\rm {u}}$ for value $\rho_{\rm 0}/\rho_{\rm u}=300$. It should be noted that the mass-to-flux estimates are not strongly dependent on the background values, which are far less than the density in the vicinity of the transonic radius.


\section{Discussion}\label{dis}
We have introduced the CFS model, a new technique to predict the magnetic field strength profile of a dense core. This model is built on a similar premise as the DCF method, where the nonthermal velocity fluctuations are assumed to be Alfv{\'e}nic. The use of $\delta B/B = \sigma_{\rm NT}/v_A$ is common to both methods. In the CFS model we measure $\delta B = \sigma_{\rm NT}(4\pi\rho)^{1/2}$, unlike the DCF method that estimates $\delta B/B$ using the dispersion ($\delta \theta$) in direction of the polarization vectors. Although similar to the DCF technique, the CFS model has a major advantage in that it predicts a field strength profile. The DCF model for a core gives only a core-average field strength estimate based on average density, average velocity dispersion and average polarization angle dispersion. For well resolved core maps in the NH$_3$ lines, the CFS model gives a finer scale predict of field structure in a core based on our choice of $\delta B/B $ at the transonic radius.
However the CFS model does not model the transition zone where there is a sharp drop of the nonthermal line width. 

The decease of the line width can be a consequence of damping of the Alfv{\'e}n waves due to reflection or dissipation across a density gradient \citep{pin12}. It could be also due to the drop in the ionization fraction at the transonic radius, leading to ambipolar diffusion damping of Alfv{\'e}n waves  .
Thus it is possible that non-ideal MHD effects may become relevant within the core.

In  Appendix \ref{appB}, we consider the effect of ambipolar diffusion on wave propagation  within the core. Equation (\ref {alf2}) gives a modified version of the Alfv{\'e}nic theory, which incorporates the correction term due to ambipolar diffusion. For conditions appropriate to a dense core, Equation (\ref{valuefxi}) shows that the use of the flux-freezing relation, Equation (\ref{alfvenic}), is approximately valid within the core. Furthermore, even though the Alfv{\'e}n waves are damped within the core, their wavelengths are long enough that they can propagate above cutoff and can be responsible for the observed nonthermal line widths.


A significant source of uncertainty in the CFS model is the value of $\beta$. As seen previously, the magnetic field strength varies by $\approx 38 \%$ when the value of $\beta$ changes from $0.5$ to $0.8$. Although results from turbulent simulations \citep[for example ][]{kud03} do constrain the value of $\beta$ to be $<1$, there is still an allowed spread in the choice of $\beta$. Another possible approach is to  assume a critical mass-to-flux ratio within the transonic radius and then derive a value of $\beta$ at the transonic radius. This yields a $\beta$ value for all cores that is close to 0.6, with the distribution having a mean and standard deviation of $0.57$ and $0.16$, respectively. Overall, the assumptions of $\beta \lesssim 1$ at the transonic radius and $ \mu \gtrsim 1 $ within the core are mutually consistent. 
Furthermore, the average interior value of $\beta \equiv \delta \theta = 0.12$ \cite{kan172} for the starless core Fest 1-457 lies within the range of estimated values of $\beta$ inside the transonic radius of H-MM1 (see Figure \ref{changeinB}).

Another source of uncertainty is in the approximation of cores as spheres in which there
is a power-law relation $B \propto \rho^{\kappa}$ for the magnetic field
strength. The cores are most likely to be spheroids that have at least some
flattening along the magnetic field direction. The relation $B \propto
\rho^{\kappa}$ actually applies to average quantities in an object that
contracts with flux freezing; $\kappa=2/3$ appropriate for spherical
contraction \citep{mes66} and $\kappa = 1/2$ appropriate for contraction with flattening
along the magnetic field direction \citep{mou76b}. The spherical model
of \citet{mes66} has features that are not present in our simplified spherical
model in which $B \propto \rho^{\kappa}$ at every interior point.
The magnetic field strength in the hourglass pattern calculated by 
\citet{mes66} is not spherically symmetric and has slightly different
profiles along the cylindrical $r$- (hereafter $x$-) and $z$- directions. 

We compared our 
Equation (\ref{magneticfieldmodel}) results to the \citet{mes66} model along both principal axes 
for clouds with central
to surface (transonic radius) density ratios of 30 and 300, and found a maximum
factor of 2 discrepancy. The values of  
the \citet{mes66} model differs most from our model along the $x$- direction,
where it can be up to a factor of 2 greater, but differs less along the $z$-
direction, where it's values are less than that of our model.
The differences decrease as the central to surface density ratio increases \citep[see also][]{mye18}. 
In the flattened magnetohydrostatic equilbrium models of \citet{mou76b} the 
magnetic field strength at the center of the cloud is about a factor of 2
less than our central value, for an equilibrium cloud with critical
mass-to-flux ratio and central to surface density 
ratio of about 20. This means that the effective value of $\kappa$ is slightly
less than $1/2$ at the center of that model. \citet{mou76b} shows in his Fig. 7 
that the central value of $\kappa$ approaches $1/2$ as greater central density 
enhancements are obtained. This can also be seen in Fig. 8 of \citet{tom88}.

In this paper, the nonthermal velocity dispersion derived from observed two-dimensional maps is used to approximate the nonthermal velocity dispersion $\sigma_{\rm NT}(r)$ as a function of spherical radius in three dimensions. This approximation overestimates $\sigma_{\rm NT}(r)$ because it treats a map of the line-of-sight average as a function of map radius as though it were a map of the nonthermal velocity dispersion along a spherical radius. The line-of-sight column density is also used to derive a density that we take to be a function of spherical radius. By comparing numbers for a Plummer sphere with $p=2$ we find that the ratio of this line-of-sight average density at a map radius to the actual density at the same value of spherical radius is about 0.75 for a wide range of $r \geq 3 r_0$ from which most of the map information is obtained

With the CFS model, we have a new tool to study the spatial profile of magnetic fields in cores with high resolution NH$_3$ line maps. Both the magnetic field strength and the hourglass morphology can be predicted from our model. See \cite{mye18} for a detailed model of hourglass morphology and comparison with a polarization map. Furthermore, the CFS model provides a prediction of the radial profile of the polarization dispersion angle, if measurable. This opens up the possibility of using high spatial resolution polarimetry maps to test the idea of Alfv{\'e}nic fluctuations in a way that is not possible with the DCF method alone.  Some progress has recently been made in this direction by \cite{kan18}, who utilize the radial distribution of the polarization angle dispersion to estimate the magnetic field strength profile in the starless core FeSt $1-457$.

\section{Conclusions}
The important results from the above study are summarized as follows.

\begin{enumerate}
\item All the observed cores in the L1688 region of the Ophiuchus molecular cloud show a sharp decrease in their nonthermal line width as they become subthermal towards the center of the core. Furthermore, in the outer part of H-MM1, H-MM2, Oph-C, and Oph-E  there is a substantial increase of $\sigma_{\rm NT}$ compared to $c_{\rm s}$. 

\item The CFS model predicts $B(r)$, the magnetic field strength as a function of radius, which we estimate to be accurate within a factor $\sim 2$. It incorporates spatially resolved observations of the nonthermal velocity dispersion $\sigma_{\rm NT}$ and the gas density in a relatively circular dense core.
\item The CFS model yields an estimate of the profile of the magnetic field fluctuations $\delta B$ and the relative field fluctuation $\delta B/ B$ inside the core. 
\item We find that the condition $\delta B/B \lesssim 1$ at the edge of the core (where $\sigma_{\rm NT}= c_{\rm s}$) is consistent with a normalized mass-to-flux ratio $\mu \gtrsim 1 $ inside the core.
\item We map the mass-to-flux ratio of cores in Ophiuchus using the CFS model. The mass-to-flux ratio is decreasing radially outward from the center of the core.
\end{enumerate}

\section*{Acknowledgements}
We thank Sarah Sadavoy, Ian Stephens, Riwaj Pokhrel, Mike Dunham and Tyler Bourke for fruitful discussions. We also thank the anonymous referee for comments that improved the presentation of results in this paper. SB is supported by a Discovery Grant from NSERC.

%
\newpage





\appendix
\section{Annularly averaged Spectra}\label{appA0}
\begin{figure}

\centering
\includegraphics[height=14 cm,width =7.2 in,trim=0mm 0mm 0mm 0mm, clip=true]{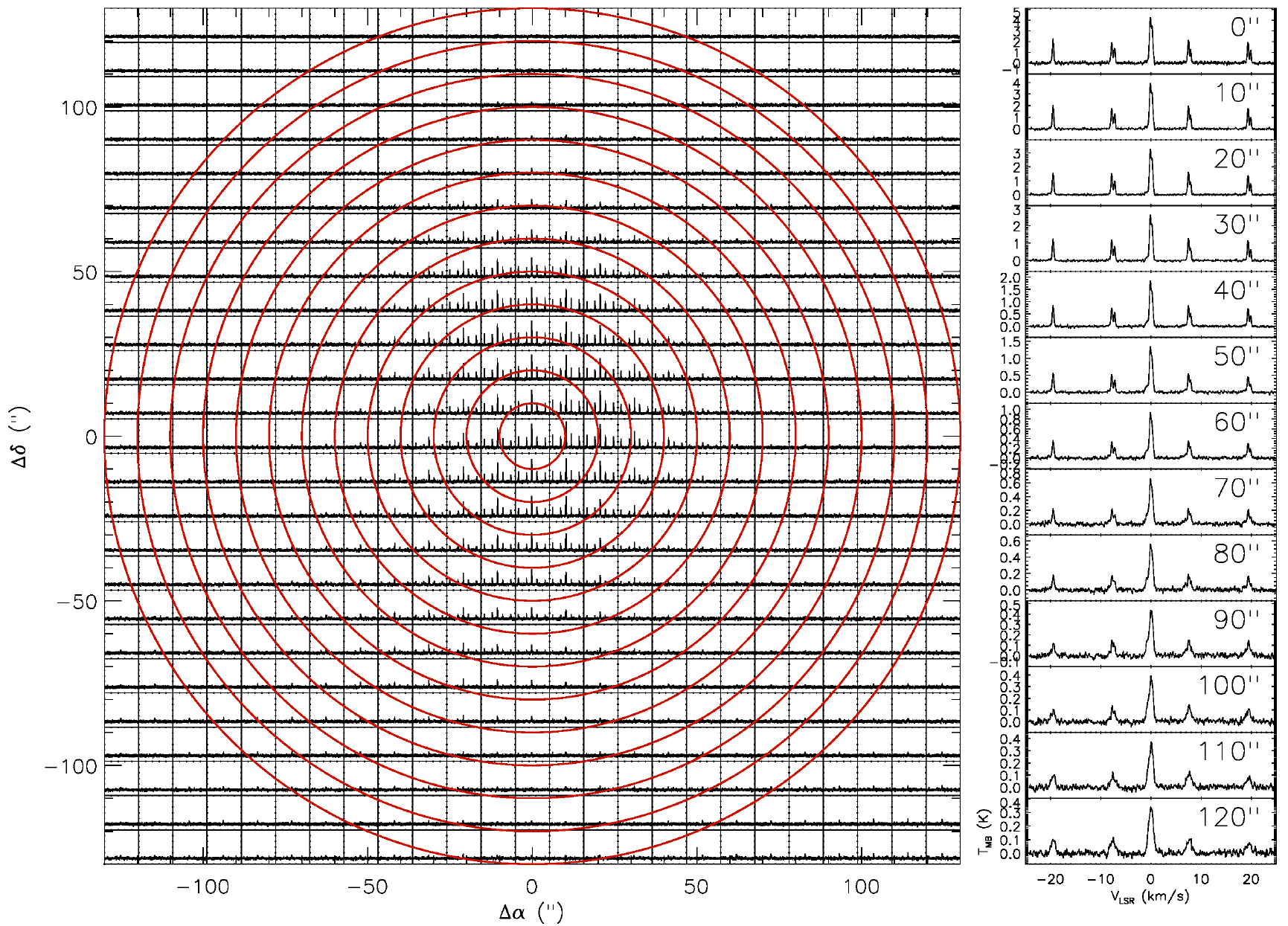}

\caption{NH$_{3}(1,1)$ spectra in a $25 \times 25$ grid around HMM-1. The averaged spectra for each of the  concentric rings are indicated on the right panel. The center of the plot corresponds to the center of the H-MM1 core. We calculate the average spectra after aligning them in velocity using the LSR velocity maps (for details see section \ref{sec:results}). The intensity of the averaged spectra decreases away from the core center.  }\label{averagedsepec}
\end{figure}

Figure \ref{averagedsepec} shows a ($25 \times 25$) grid of NH$_{3}(1,1)$ spectra around HMM-1.  The center of the plot corresponds to the center of the H-MM1 core. For each concentric ring we calculate the average spectra after aligning them in velocity using the LSR velocity maps. The averaged spectra for each of the rings are shown on the right panel. We apply the same procedure to obtain the averaged NH$_{3}(1,1)$ and NH$_{3}(2,2)$  spectra for all the cores studied in this paper. These annularly averaged spectra are then used to extract the thermal and nonthermal components of the velocity dispersion, as described in Section \ref{sec:results}. 

\section{Mass and Flux of a cylindrical tube}\label{appA}

\begin{figure}
\centering

\includegraphics[height=3.5in,width=3.5in,trim=0mm 0mm 0mm 0mm, clip=true]{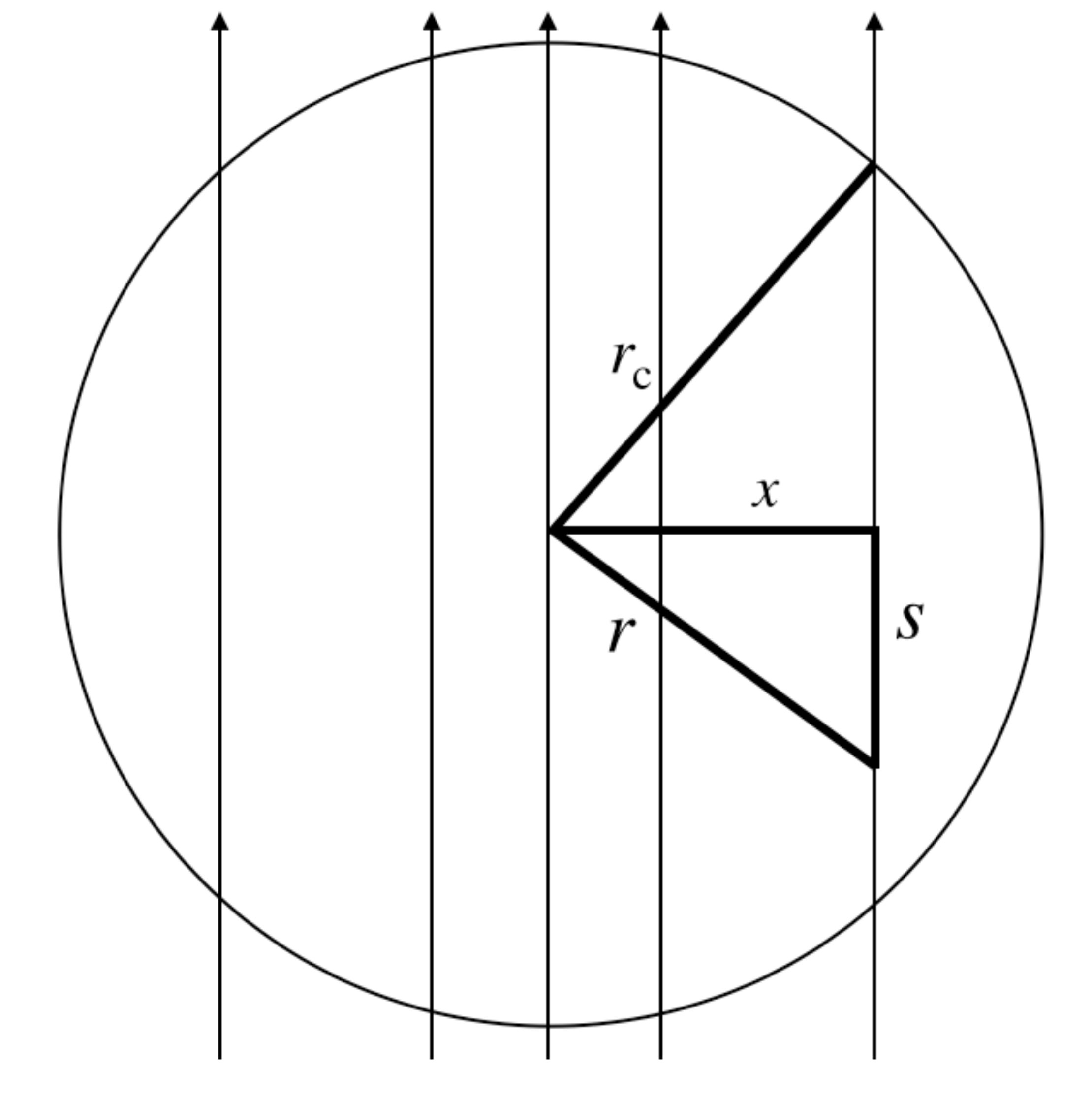}
\caption{Schematic illustration of a cut through a spherical core of radius $r_{\rm c}$. The vertical arrows represent the magnetic field lines in the plane of the sky. The  column density $\Sigma$ as a function of the offset $x$ is obtained by integrating along the direction $s$ parallel to the magnetic field lines.}\label{core_model}
\end{figure}

Here we consider a simple case where the magnetic field lines are assumed to be vertically threading a spherical core in the plane of the sky. We calculate the enclosed mass within cylindrical tubes of constant magnetic field strength. 
We integrate the volume density $\rho(r)$ given in Equation (\ref{plummer}) with $p/2 =1$ along the magnetic field lines (assumed to be vertical). See Figure \ref{core_model} for a schematic of the integration. The column density is
\begin{eqnarray}
\Sigma(x) =& 2  \int^{\sqrt{r_{\rm c}^{2} -x^2}}_0 \rho (s) ds \\ \nonumber
=& 2 \int^{r_{\rm c}}_x \frac{\rho (r)r dr}{\sqrt{r^2-x^2}},
\end{eqnarray}
where $x$ is the offset from the center in the midplane (see \cite{dap09} for an analogous calculation). The column density is then 
\begin{equation} \label{sigma}
\Sigma(x)= \frac{2 r_{0}^{2}\rho_{0}}{\sqrt{(x^{2}+r_{0}^{2})}}\arctan\left(\frac{\sqrt{r_{c}-x^{2}}}{\sqrt{r_{0}+x^{2}}}  \right).
\end{equation} 
We find the mass of the cylindrical tubes by integrating the column density from the center to a given distance $x$ in the midplane: 

\begin {equation} \label{mass}
M(x) = 2 \pi \int^{x}_{0} x^{\prime} \Sigma(x^{\prime}) dx^{\prime} .
\end {equation}
Inserting Equation (\ref{sigma}) in Equation (\ref{mass}) and integrating, we find 

\begin{eqnarray}
M(x) = 4 \pi \rho_{0} r_{0}^{2} \bigg[ r_{c} - r_{0}\, \arctan \frac{r_{c}}{r_{0}} - \sqrt{r_{c}^{2} - x^{2}}&  \\  \nonumber
 +\sqrt{r_{0}^{2} - x^{2}} \arctan \left( \frac{\sqrt{r_{c}-x^{2}}}{\sqrt{r_{0}+x^{2}}}  \right)& \bigg].
\end{eqnarray}
The corresponding magnetic flux is estimated by integrating the magnetic field profile in the horizontal midplane of the core:
\begin{equation}\label{fluxequation}
\Phi(x) = 2\pi \int^{x}_{0} B \, x^{\prime} dx^{\prime} . 
\end{equation}
Using Equation (\ref{magneticfieldmodel}) and Equation (\ref{plummer}) in the above equation yields
\begin{equation}
\Phi(x) = 2 \pi B_{c} r_{0}^2 \left(\frac{n_0}{n_{c}}\right)^{\kappa} \left[ \frac{(1 + (x/r_0)^2)^{(1-\kappa)}}{2-2\kappa} - \frac{1}{2-2\kappa}\right].
\end{equation}

For the more general case where $p/2 \ne 1$, we can use numerical integration to estimate the mass and the flux of a given core. 

\section{Dispersion relation}\label{appB}
The dispersion relation of Alfv{\'e}n waves in a partially ionized medium \citep{pin12} in the long wavelength limit (i.e., $\lambda \gg 2\pi v_{\rm A} \tau_{\rm ni} $) is 

\begin{equation}\label{DR}
\omega^{2} - k^2 v_{\rm A}^2 + i \eta_{\rm AD}  k^2 \omega  = 0.
\end{equation}
Here $\eta_{\rm AD} = v_{\rm A}^{2} \tau_{\rm ni} $ is the ambipolar diffusion resistivity and $\tau_{\rm ni} = (\gamma_{\rm ni} \rho_{\rm i})^{-1}$ is the mean neutral-ion collision time in terms of the drag coefficient 
\begin{equation}\label{tau}
\gamma_{\rm ni} = \frac{\langle \sigma w\rangle_{\rm in}}{ 1.4\,(m_{n} + m_{i})}
\end{equation}
\citep{bas94}, and the ion density $\rho_{\rm i}$. In the above equation $\langle \sigma w\rangle_{\rm in}$ is the average collision rate between the ions of mass $m_{\rm i}$ and neutrals of mass $m_{\rm n}$.
On rearranging, Equation (\ref{DR}) is rewritten as
\begin{equation}\label{DR2}
k^2 = \frac{\omega^2}{v_{\rm A}^2} \left[\frac{1}{1+ \textit{i}\omega \tau_{\rm ni} } \right].
\end{equation}
In the limit $\omega \tau_{\rm ni}  < 1$, Equation (\ref{DR2}) on binomial expansion yields
\begin{equation}\label{D3}
k = \frac{\omega}{v_{\rm A}} (1 - \textit{i}\frac{1}{2}\omega \tau_{\rm ni}).
\end{equation}
Defining $\xi = (1/2) \omega \tau_{\rm ni} $ as a dimensionless parameter, we can represent Equation (\ref{D3}) in terms of a magnitude and a phase $\theta$: 
\begin{equation}
\left| \frac{k}{\omega} \right|  = \frac{1}{v_{\rm A}} \sqrt{1 + \xi^{2}}\left| e^{\textit{i} \theta} \right| .
\end{equation}
Using Equation (\ref{D3}) to replace $k$ in Equation (17) from \cite{pin12}, we derive the relation between the amplitude of fluctuation of the neutral velocity $u_{\rm n0}$  to the  fluctuation of the magnetic field $\delta B$. 
In the long wavelength limit we get 

\begin{equation}\label{alf1}
|u_{\rm n0}| = v_{\rm A} \frac{|\delta B|}{B} \sqrt{1+ \xi^{2}}.
\end{equation}
If $\xi \ll 1$, then 

\begin{equation}\label{alf2}
|u_{\rm n0}| \simeq v_{\rm A} \frac{|\delta B|}{B}\left [1 + \frac{1}{2} \xi^{2}\right ].
\end{equation}
This gives a modified version for the Alfv{\'e}nic theory, which incorporates the correction term due to ambipolar diffusion. Equation (\ref{alf2}) is equivalent to Equation (\ref{alfvenic}), if we equate $\sigma_{\rm NT} = |u_{\rm {n0}}| $ and take the limit $\xi \rightarrow 0$.
Again assuming that $\xi\ll 1$ (which we will later verify), we apply the standard dispersion relation of ideal Alfv{\'e}n waves, $\omega = v_{\rm A} k $, and express $\xi$ in terms of wavenumber $k$:,

\begin{equation} \label{xi}
\xi^{2} = \left(\frac{v_{\rm A }\tau_{\rm ni} k}{2} \right)^{2}.
\end{equation}
Using Equation (\ref{tau}) to replace $ \tau_{\rm ni}$ in terms of the drag coefficient $\gamma_{\rm {ni}}$ and ion density,  we get 
\begin{equation}
\xi^{2} = \frac{1}{4}\frac{B^{2}}{4 \pi \rho_{\rm n}} k^{2} \left[ 1.4 \, \frac{m_{\rm i}+ m_{\rm n}}{\rho_{\rm i} \langle \sigma w\rangle_{\rm in}}\right]^{2}.
\end{equation}
We can estimate the above quantities in Equation (\ref{xi}) by specifying appropriate values relevant for dense cores embedded in molecular clouds. For example, if   $B \simeq30 \ \mu G$ and $\rho_{n} = m_{\rm n} n_{0}$, where $n_{0} = 10^{4}\, \rm {cm^{-3}}$ is the number density of neutrals and $m_{\rm n} =2.33 \times m_{\rm H} $, the Alfv{\'e}n speed $v_{\rm A} = 0.4 \,\rm {km \,s^{-1}}$. 
Furthermore, for $\langle \sigma w\rangle_{\rm in} = 1.69 \times 10^{-9} \rm \, {cm^{-3}} \rm {s^{-1}} $ and $m_{\rm i} = 29 \times m_{\rm H}$, the drag coefficient $\gamma_{\rm ni} = 2.3 \times 10^{13} \, \rm {cm ^{3} g^{-1} s^{-1}}$. The ion density $\rho_{\rm i}$ is determined by the approximate relation
\begin{equation}
\rho_{\rm i} = m_{\rm i} K n_{0}^{1/2} = 1.45\times 10^{-23} \rm {g\, cm^{-3}},
\end{equation}
where $K = 3 \times 10^{-3} \, \rm {cm^{-3} }$ \citep{elm79}. This gives $ \tau_{\rm ni} = (\gamma_{\rm ni} \rho_{\rm i})^{-1} =3 \times 10^{9} \, \rm {s}$.    
The wavenumber $k = (2 \pi) /\lambda$ of interest will roughly correspond to a wavelength $\lambda \simeq 0.1 \, \rm {pc}$, i.e., about equal to the core diameter. This yields 

\begin{equation}\label{valuefxi}
\xi = 1.3 \times 10^{-3}  \left(\frac{v_{\rm A}}{ 0.4 \, \rm {km \, s^{-1}}}\right) \left( \frac{\tau_{\rm ni}}{ 3 \times 10^{9 } \rm {s}}\right) \left(\frac{0.1\, \rm {pc }}{\lambda}\right) ,
\end{equation}
such that $\xi \ll 1$ (justifying the approximation we made in eq. B13) in a dense core. This is equivalent to the waves having wavelength $\lambda \gg  \lambda_{\rm cr}$, where $\lambda_{\rm cr} = \pi v_{\rm A} \tau_{\rm ni}$, (see Eq. (15) in \cite{pin12}) is the critical wavelength for wave propagation, i.e., wavelengths shorter than $\lambda_{\rm {cr}}$ are critically damped. Thus, the Alfv{\'e}n waves can still propagate within the core and be responsible for the observed nonthermal line widths. The condition $\xi \ll 1$ continues to apply for wavelengths $\lambda$ significantly smaller that $0.1 \rm pc$, as can be seen from Equation (\ref {valuefxi}).

\end{document}